\documentclass[%
reprint,
 amsmath,amssymb,
 aps,
 pra,
]{revtex4-2}

\newcommand{\env}[1]{\texttt{#1}}
\usepackage{graphicx}% Include figure files
\usepackage{dcolumn}% Align table columns on decimal point
\usepackage{bm}% bold math
\usepackage{floatrow}
\usepackage{amsmath}
\usepackage{comment}
\usepackage{mathrsfs}
\usepackage{xcolor}
\usepackage{hyperref}
\usepackage{cleveref}
\crefname{section}{§}{§§}
\Crefname{section}{§}{§§}
\hypersetup{
  colorlinks   = true, %Colours links instead of ugly boxes
  urlcolor     = blue, %Colour for external hyperlinks
  linkcolor    = blue, %Colour of internal links
  citecolor   = blue %Colour of citations
}

\begin{document}

\preprint{APS/123-QED}

\title{Dynamics of coupled thermoacoustic oscillators\\ under asymmetric forcing: Experiments and theoretical modeling}% Force line breaks with \\
% \thanks{A footnote to the article title}%

\author{Ankit Sahay}\email{ankitsahay02@gmail.com}\affiliation{Department of Aerospace Engineering, Indian Institute of Technology Madras, Chennai 600036, India}\author{Amitesh Roy}\affiliation{Department of Aerospace Engineering, Indian Institute of Technology Madras, Chennai 600036, India}\author{Samadhan A. Pawar}\affiliation{Department of Aerospace Engineering, Indian Institute of Technology Madras, Chennai 600036, India}\author{R. I. Sujith}\affiliation{Department of Aerospace Engineering, Indian Institute of Technology Madras, Chennai 600036, India}

%\author{Ankit Sahay, Amitesh Roy, Samadhan A. Pawar, R. I. Sujith}
% \altaffiliation[Also at ]{Physics Department, XYZ University.}%Lines break automatically or can be forced with \\
% \author{Ankit Sahay}%
% \email{ankitsahay02@gmail.com}
% \affiliation{Department of Aerospace Engineering, Indian Institute of Technology Madras, Chennai 600036, India}

\date{\today}% It is always \today, today,
             %  but any date may be explicitly specified
\begin{abstract}
Quenching of limit cycle oscillations (LCO), either through mutual coupling or external forcing, has attracted wide attention in several fields of science and engineering. However, the simultaneous utilization of these coupling schemes in quenching of LCO has rarely been studied despite its practical applicability. We study the dynamics of two thermoacoustic oscillators simultaneously subjected to mutual coupling and asymmetric external forcing through experiments and theoretical modeling. We investigate the forced response of both identical and non-identical thermoacoustic oscillators for two different amplitudes of LCO. Under mutual coupling alone, identical thermoacoustic oscillators display the occurrence of partial amplitude death and amplitude death, whereas under forcing alone, asynchronous quenching of LCO is observed at non-resonant conditions. When the oscillators are simultaneously subjected to mutual coupling and asymmetric forcing, we observe a larger parametric region of oscillation quenching than when the two mechanisms are utilized individually. This enhancement in the region of oscillation quenching is due to the complementary effect of amplitude death and asynchronous quenching. However, a forced response of coupled non-identical oscillators shows that the effect of forcing is insignificant on synchronization and quenching of oscillations in the oscillator which is not directly forced. Finally, we qualitatively capture the experimental results using a reduced-order theoretical model of two coupled Rijke tubes which are coupled through dissipative and time-delay coupling and asymmetrically forced. We believe that these findings offer fresh insights into the combined effects of mutual and forced synchronization in a system of coupled nonlinear oscillators.
\end{abstract}
%\keywords{Suggested keywords}%Use showkeys class option if keyword
                              %display desired
\maketitle

%\tableofcontents

\section{\label{Introduction} Introduction}
Coupled interacting nonlinear oscillators appear extensively in physical systems around us \cite{jenkins2013self}. Depending upon the nature of coupling, a population of interacting oscillators can synchronize starting from an initially incoherent or desynchronized state. Such synchronized oscillators can exhibit an emergence of ordered patterns as observed in several examples in nature, such as flocking of bird or flashing of light by fireflies \cite{strogatz2004sync}. However, synchronization of oscillations may be detrimental sometimes, and may need to be desynchronized \cite{strogatz2005crowd} or quenched \cite{biwa2015amplitude}. Here, we study the practical application of synchronization theory in attaining control of self-excited oscillations in a system of coupled prototypical thermoacoustic oscillators. A thermoacoustic oscillator refers to a confined combustion system (for example, gas turbine combustor and rocket engines). In such a system, positive feedback between the acoustic pressure oscillations and the heat release rate oscillations leads to the generation of large amplitude, self-sustained periodic oscillations in the acoustic field. The occurrence of these large amplitude oscillations is referred to as thermoacoustic instability (TAI). The presence of such instabilities can inflict considerable damage to mechanical and structural components used in gas turbine and rocket engines \cite{lieuwen2005combustion, sujith2020complex, pikovsky2003synchronization}. Complex interactions between the acoustic field, turbulent flow field, and heat release rate field has recently led to the widespread use of complex systems approach to understand the phenomenon of thermoacoustic instability \cite{sujith2020complex, juniper2018sensitivity}.

In general, there are two types of interactions leading to synchronization of oscillators: mutual and forced \cite{pikovsky2003synchronization}. In the former, oscillators mutually interact through bidirectional coupling, leading to the adjustment of phases and frequencies of both the oscillators to a common state of mutual synchronization. Oscillation quenching attained through the phenomenon of amplitude death (AD) in mutually coupled systems is an exciting prospect that has been shown to work for various systems \cite{pyragas2000stabilization, kumar2011role, huddy2012amplitude, mirollo1990amplitude}. Amplitude death in a population of strongly coupled oscillators refers to a situation where all the oscillators pull each other off of their oscillatory state into the same stable fixed point, leading to complete cessation of oscillations. If the mutual coupling is not strong enough, the situation of partial amplitude death (PAD) may arise, where some oscillators retain their oscillatory behavior while oscillations are ceased in others \cite{atay2003total}. On the other hand, in the case of forced synchronization, an independent master system (an external force) drives a slave system (driven oscillator); thus, forming a unidirectional master-slave system \cite{balanov2008synchronization}. In this type of coupling, the driven system adjusts its phase and frequency to that of the external forcing during the state of synchronization. Depending upon the frequency and the amplitude of forcing, the natural oscillations can be phase-locked to forcing, and in some cases, the self-excited oscillations can be completely suppressed through the phenomenon of asynchronous quenching  \cite{keen1970suppression, mondal2019forced}.

Quite a few numerical studies have been conducted to investigate the simultaneous effect of forcing in a system of coupled oscillators. In a system of two weakly coupled Van der Pol oscillators, Battelino \cite{battelino1988persistence} observed that when each oscillator is externally forced, and a constant phase difference is present between the forcing signals applied to each of the oscillators, the system exhibits three-frequency quasiperiodicity. As the coupling strength and forcing amplitude are increased, three-frequency quasiperiodicity is first replaced by a two-frequency quasiperiodic regime, and subsequently by phase-locked periodic and chaotic regimes. The transition between the phase-locked region and the two- and three-frequency quasiperiodic regions takes place through saddle-node bifurcations \cite{anishchenko2009numerical, anishchenko2009phase}. The external force first destroys the regime of mutual synchronization of oscillators. As the forcing amplitude is increased, the oscillator which is subjected to external forcing synchronizes with the forcing signal first, followed by the forced synchronization of the entire system \cite{anishchenko2008bifurcational}. Similar observations were made in a system of coupled Van der Pol and Duffing oscillators \cite{wei2011nonlinear}. Such an approach was utilized for modeling the effect of a pacemaker on the dynamics of the human heart \cite{honerkamp1983heart}. On the flip side, to the best of our knowledge, only one experimental study has been conducted to understand the effect of forcing on mutually coupled electronic circuits with the aim of verifying the phase dynamics associated during regimes of desynchronized, mutually synchronized, two-frequency quasiperiodic, three-frequency quasiperiodic, and forced synchronized oscillations \cite{anishchenko2009numerical}. Further, all the aforementioned studies focus on the phase dynamics of the system with particular attention afforded to the dynamics of bifurcations among quasiperiodic, periodic, and chaotic regimes. In light of the above discussed limitations of past studies, there is a need to experimentally quantify the effect of mutual coupling and external forcing on the amplitude of the limit cycle oscillations (LCO) observed in both the oscillators. Thus, we analyze the phase as well as the amplitude response of two coupled thermoacoustic oscillators under the condition of asymmetric external forcing, and model the observations with a low-order model. This makes up the key objective of the present study, and is of general interest to the nonlinear dynamics community.

Biwa \textit{et al.} \cite{biwa2015amplitude} were the first to experimentally demonstrate the control of LCO through amplitude death in coupled thermoacoustic engines using both the dissipative and time-delay couplings. Thomas \textit{et al.} \cite{thomas2018effect} investigated amplitude death in a model of coupled Rijke tubes. They found that the simultaneous presence of dissipative and time-delay couplings was far more effective in attaining AD in the Rijke tubes than either of the two types of coupling applied separately. In a follow-up study with additive Gaussian white noise, Thomas \textit{et al.} \cite{thomas2018noiseeffect} observed that the presence of noise affects the coupled behavior of oscillators and mutual coupling only leads to about 80\% suppression in the amplitude of the uncoupled LCO as opposed to the state of AD observed in the absence of noise in the system. Dange \textit{et al.} \cite{dange2019oscillation} performed detailed experimental characterization of coupled Rijke tube oscillators and found that only time delay coupling is sufficient to achieve AD of low amplitude LCO. In oscillators undergoing high amplitude oscillations, frequency detuning is needed in addition to time delay coupling for attaining AD and PAD. They also reported the phenomenon of phase-flip bifurcation in the coupling of identical thermoacoustic oscillators. Recently, Hyodo \textit{et al.} \cite{hyodo2020suppression} used double-tube coupling method to significantly reduce the tube diameter necessary for demonstrating AD in a system of coupled identical Rijke tubes. Jegal \textit{et al.} \cite{jegal2019mutual} demonstrated the occurrence of AD in a practical turbulent system consisting of two lean-premixed model swirl combustors. Moon \textit{et al.} \cite{moon2020mutual} compared the characteristics of mutual synchronization of these systems when the length and diameter of the coupling tube were changed. Jegal \textit{et al.} \cite{jegal2019mutual} also found that under different conditions, mutual synchronization can also trigger the strong excitation of a new mode, even when the two combustors are individually stable in the absence of coupling.

On the other hand, Balusamy \textit{et al.} \cite{balusamy2015nonlinear} experimentally studied the forced response of LCO in a swirl-stabilized turbulent combustor using the framework of forced synchronization and observed different states such as phase-locking, phase drifting, and phase trapping in the system. Kashinath \textit{et al.} \cite{kashinath2018forced} highlighted the route to forced synchronization of limit cycle, quasiperiodic, and chaotic oscillations in a numerical model of the premixed flame. These findings were verified in experiments on Rijke tubes \cite{mondal2019forced} and laminar combustors \cite{guan2019open, roy2020mechanism, guan2019forced}.  Mondal \textit{et al.} \cite{mondal2019forced}, Guan \textit{et al.} \cite{guan2019open} and Roy \textit{et al.} \cite{roy2020mechanism} further showed that forcing at frequencies  away from the natural frequency of LCO lead to greater than 80\% decrease in their amplitude through a phenomenon known as asynchronous quenching. 

The aforementioned studies demonstrate the possibility of controlling thermoacoustic instability based on mutual or forced synchronization. However, the scope of these studies is still limited as external forcing can quench thermoacoustic instability in a single system in a specific range of forcing parameters, whereas practical engines generally have multiple combustors working in tandem. Hence, the information on how the forcing of thermoacoustic instability in one combustor affects the thermoacoustic instability developed in another combustor is still unknown. Similarly, mutual coupling works for two coupled oscillators; however, the parametric regime for which amplitude death is observed is limited. Therefore, there is a need to combine both these methodologies to overcome their individual limitations. Essentially, we aim to implement a proof-of-concept capable of combining the best of both strategies – asynchronous quenching and mutual synchronization – to control thermoacoustic oscillations. Towards this purpose, we couple two Rijke tubes during the state of thermoacoustic instability and subject one to external harmonic forcing (asymmetric forcing). We then measure the phase and amplitude response of the resultant acoustic pressure oscillations in the combustors at different conditions of forcing and coupling parameters. We perform asymmetric forcing experiments on identical and non-identical thermoacoustic oscillators to comprehensively assess the response to forcing. We find that through asymmetric forcing, we can expand the region of oscillation quenching of thermoacoustic instability in the system of coupled identical Rijke tubes by compounding the effect of asynchronous quenching and mutual synchronization. Finally, we develop a model where two Rijke tube oscillators are coupled through dissipative and time-delay coupling and are forced asymmetrically. We show that the model compares favorably with the experimental results, indicating the usefulness of reduced-order modeling of coupled oscillator models under forcing.

The rest of the paper is organized as follows. In \S\ref{2. Experimental Setup} we describe the experimental setup. In \S\ref{Results:1}, we investigate the forced synchronization characteristics of a single Rijke tube. We then demonstrate the presence of AD and PAD in mutually coupled Rijke tubes in \S\ref{Results:2}. In \S\ref{Results:3}, we characterize the response of coupled identical Rijke tubes to asymmetric forcing. We consider the response of non-identical oscillators in \S\ref{Results:4}. In \S{\color{blue}IV}, we describe the model and numerically investigate the dynamics of coupled Rijke tube oscillators under forcing. Finally, we present the conclusions from the study in \S\ref{Conclusions}.

% In \cref{Setup}, we introduce the experimental setup. In \cref{Results:1}, we investigate the forced synchronization characteristics of a single Rijke tube. We then demonstrate the presence of AD and PAD in mutually coupled Rijke tubes in \cref{Results:2}. In \cref{Results:4}, we characterize the response of coupled identical Rijke tubes to asymmetric forcing. We consider the response of non-identical oscillators in \cref{Results:6}. Finally, we present the conclusions from the study in  \cref{Conclusion}.

\section{Experimental Setup}
\label{2. Experimental Setup}

\begin{figure}[t!]
\centering
\includegraphics[width=1\textwidth]{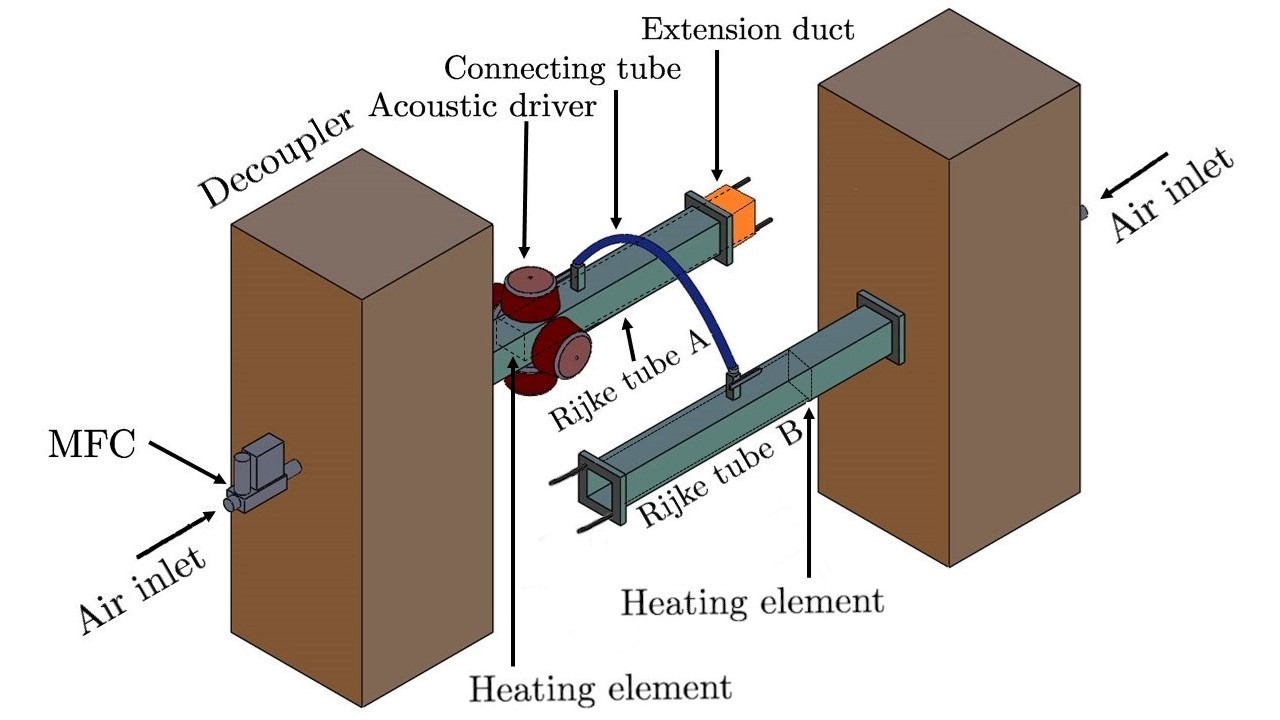}
\caption{\label{setup} Schematic of the experimental setup having two horizontal Rijke tube oscillators A and B, which are coupled using a connecting tube. Rijke tube A is acoustically forced with 4 acoustic drivers attached to its sides. An extension duct in Rijke tube A is used for implementing frequency detuning in the system.}
\end{figure}

The experimental setup used in this study consists of a pair of horizontal Rijke tubes (Fig. \ref{setup}). Rijke tube A has a cross-section of $ 9.3 \times 9.4$ $\text{cm}^2$  with a length of $102$ cm. Rijke tube B has a cross-section of $ 9.3 \times 9.5$ $\text{cm}^2$ and a length of $104$ cm. A decoupler of dimensions $ 102 \times 45 \times 45$ $\text{cm}^3$ is attached to the inlet of the respective Rijke tube to eliminate upstream disturbances and ensure a steady flow in the system. Each Rijke tube consists of a separate electrically heated wire mesh, powered by an external DC power supply, which acts as a compact heat source. The heaters are located $27$ cm downstream of the inlet in each duct. The air flow rate is maintained constant in each of the Rijke tubes through separate mass flow controllers (MFC, Alicat Scientific) of uncertainty $\pm$(0.8\% of the measured reading + 0.2\% of the full scale reading).

Both the Rijke tubes are coupled using a single vinyl tube of length $L$ and internal diameter $d$ (see Fig. \ref{setup}). The ports for the connecting tube are located $57$ cm downstream of the inlet. Ball-type valves are manually opened to establish coupling between the two oscillators. Four wall mounted acoustic drivers (Minsound TD-200A) are attached on each side of Rijke tube A, $41$ cm downstream of the inlet. The acoustic drivers are connected in parallel to a power amplifier (Ahuja UBA-500M). Sinusoidal forcing signal with mean-to-peak amplitude ($A_f$, in mV) and frequency ($f_f$, in Hz), generated using a Tektronix function generator (Model No. AFG1022), is input to the power amplifier to drive the coupled system.

A piezoelectric pressure transducer (PCB 103B02, sensitivity $217.5$ mV/$\text{kPa}$, uncertainty $\pm 0.15$ Pa) is mounted close to the midpoint along the length of the duct in each Rijke tube and is used for measuring the acoustic pressure fluctuations in the system. Data are acquired simultaneously from both the Rijke tubes at a sampling rate of $10$ kHz for a duration of $5$ s for each parametric condition using a data acquisition system (NI USB 6343). The resolution of frequency in the power spectrum of the signals is equal to 0.2 Hz. The rate of decay of an acoustic pulse generated from loudspeakers in the absence of flow is used to measure the damping in the two Rijke tubes. The acoustic decay rate values for Rijke tubes A and B are measured to be $16.5 \pm 2$ $\text{s}^{-1}$ and $12.9 \pm 1.8$ $\text{s}^{-1}$, respectively. To ensure repeatability of the results and consistency of the experimental conditions, the experiments are conducted only when the measured acoustic decay rates lie in the range mentioned above. A change in the decay rate only changes the critical parameter values at which the different dynamical states are observed; however, the overall dynamics remain the same. During each experiment, both the Rijke tubes are preheated for $10$ minutes by supplying DC power at $1$ V to the wire mesh. The preheating ensures a steady temperature profile inside the Rijke tubes.

\begin{figure*}[t!]
\centering
\includegraphics[width=0.75\textwidth]{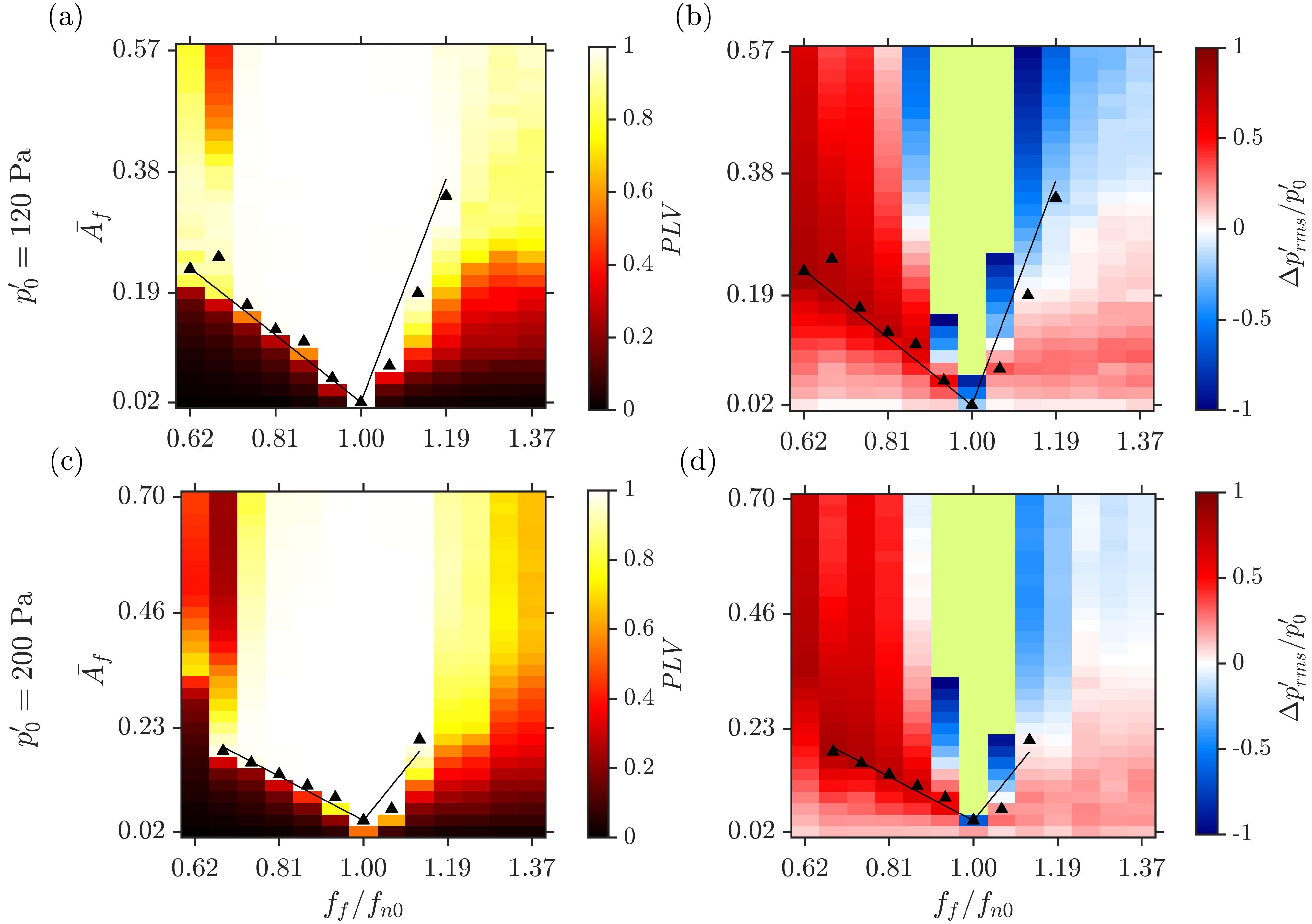}
\caption{\label{single_osc}Forced response of a single Rijke tube. (a,c) The phase response shown in terms of $PLV$ and (b,d) amplitude response in terms of fractional change in the amplitude of LCO in the Rijke tube for (a,b) $p^\prime_{0} = 120$ Pa and (c,d) $p^\prime_{0} = 200$ Pa for different values of $\bar{A_f}$ and $f_f$. The synchronization boundaries in (a-d) are obtained through least-square-fit of points where $PLV=0.98$. The $R^2$ values for the least-square fitting are given in Section I of the Supplemental Material. The region of forced synchronization decreases with increase in $p^\prime_{0}$ (a,c), while quenching of LCO is observed only for $f_f<f_{n0}$ (b,d). Green region  around $f_f/f_{n0} = 1$ in (b,d) signifies the amplification of LCO above twice the value of unforced amplitude, such that $\Delta p^\prime_{rms}/p^\prime_{0}$ varies in the range (-5.76, -1) in (b), and (-4.47, -1) in (d).}
\end{figure*}

A telescopic slide mechanism of 12 cm length is used to vary the length of Rijke tube A. The natural frequency of Rijke tube A can be varied between $f_{n0} = 162 $ Hz to $f_{n0} = 148 $ Hz with an uncertainty of $\pm 2$ Hz due to the uncertainty in measuring the length of the Rijke tube. After preheating both the Rijke tubes, the heater power is increased such that the system undergoes subcritical Hopf bifurcation. The heater power is maintained away from the bistable region in all experiments. LCO are maintained in each of the Rijke tubes before coupling is induced. The amplitude of the LCO are controlled by varying the heater power and the air flow rate. All the experiments are reported for two different amplitudes (root-mean-square) of the LCO in both the Rijke tubes in the uncoupled state: (a) lower amplitude $p^\prime_{0} = 120$ Pa maintained by supplying a constant air flow rate of $2.4 \pm 0.01$ g/s (Re = $1370$) in each Rijke tube, and (b) higher amplitude $p^\prime_{0} = 200$ Pa maintained by supplying a constant air flow rate of $3.95 \pm 0.01$ g/s (Re = $2284$) in each Rijke tube. After both the oscillators exhibit LCO in their uncoupled state, the valves are opened to couple these oscillators. The coupled system is then asymmetrically forced through the loudspeakers connected to Rijke tube A.

\section{Results and discussions}
\label{3. Results and discussions}

\subsection{Response of a thermoacoustic oscillator to external forcing}
\label{Results:1}
We begin by investigating the forced acoustic response of  LCO in a single Rijke tube. Figure \ref{single_osc} shows a two-parameter bifurcation plot on an $\bar{A_f}-f_f$ plane illustrating the phase (Figs. \ref{single_osc}a,c) and amplitude (Figs. \ref{single_osc}b,d) response of LCO under external forcing. The natural frequency of LCO is $f_{n0}=160 \pm 2$ Hz. The forced response is studied for two different amplitudes of unforced LCO: $p^\prime_{0}=120$ Pa corresponding to $p_{0}=26$ mV (Figs. \ref{single_osc}a,b) and $p^\prime_{0}=200$ Pa corresponding to $p_{0}=43$ mV (Figs. \ref{single_osc}c,d). Here, $p^\prime_{0}$ refers to the amplitude of the unforced LCO in pascal (Pa), and $p_{0}$ refers to the equivalent reading obtained from the piezoelectric transducer in mV. The values of forcing amplitude  $A_f$ (in mV) are normalized with the unforced amplitude of LCO measured in mV such that $\bar{A_f} = A_f/p_0$.

In Figs. \ref{single_osc}a,c, we plot the distribution of phase locking value ($PLV$) between the forced LCO and the external forcing signal on the $\bar{A_f}-f_f$ plane. $PLV$ quantifies the degree of synchronization between a pair of oscillators at any given condition of forcing ($A_f$, $f_f$). It is defined as \cite{pikovsky2003synchronization},
\begin{equation}
    PLV = \dfrac{1}{N} \Bigg{\lvert} \sum\limits_{n=1}^N \text{exp}(i \Delta \phi) \Bigg{\lvert},
\end{equation}
where $N$ is the length of the $p^\prime$ signal, and $\Delta \phi$ is the instantaneous phase difference between the $p^\prime$ and forcing signals. Here, forcing is assumed to be sinusoidal of the form $F(t) = A_f \hspace{3 pt} \text{sin}(2\pi \hspace{1 pt} f_f \hspace{1 pt} t)$. The instantaneous phase of the signals is determined using the analytic signal approach utilizing the Hilbert transform \cite{pikovsky2003synchronization}. A $PLV$ of $1$ indicates synchronization of two oscillators, while a $PLV$ of $0$ indicates desynchronization of the oscillators. In experimental situations, it is difficult to obtain $PLV=1$; hence, we denote phase synchronization boundaries of LCO through a least-square-fit of the points where $PLV=0.98$. The $R^2$ values of the least-square-fit lines are given in Section I of the Supplemental Material. The V-shaped region, called the Arnold tongue, separates the region of forced synchronization from the region desynchronization. The critical amplitude of forcing required for forced synchronization of LCO, $A_f = A_{f,c}$, increases almost linearly with an increase in the frequency difference $ \Delta f = \vert f_f-f_{n0} \vert$.

From Figs. \ref{single_osc}a,c, we notice that for $p^\prime_{0}=200$ Pa, the range of $f_f$ for which forced synchronization occurs in the system is smaller than that observed for $p^\prime_0=120$ Pa, indicating the dependence of forced synchronization of LCO on their amplitude in the unforced state. Further, at higher values of $\Delta f$, achieving forced synchronization of both LCO becomes difficult. We also observe regions of desynchronized oscillations (indicated in red) in the range of $f_f/f_{n0} = 0.62-0.70$ at high values of $\bar{A_f}$ ($\bar{A_{f}}> 0.35$ in Figs. \ref{single_osc}a,c). The reason for desynchrony in this region is a result of period-3 behavior in $p^\prime$, which leads to low values of \textit{PLV} (see Section II of the Supplemental Material).

\begin{figure*}[t!]
\centering
\includegraphics[width=0.80\textwidth]{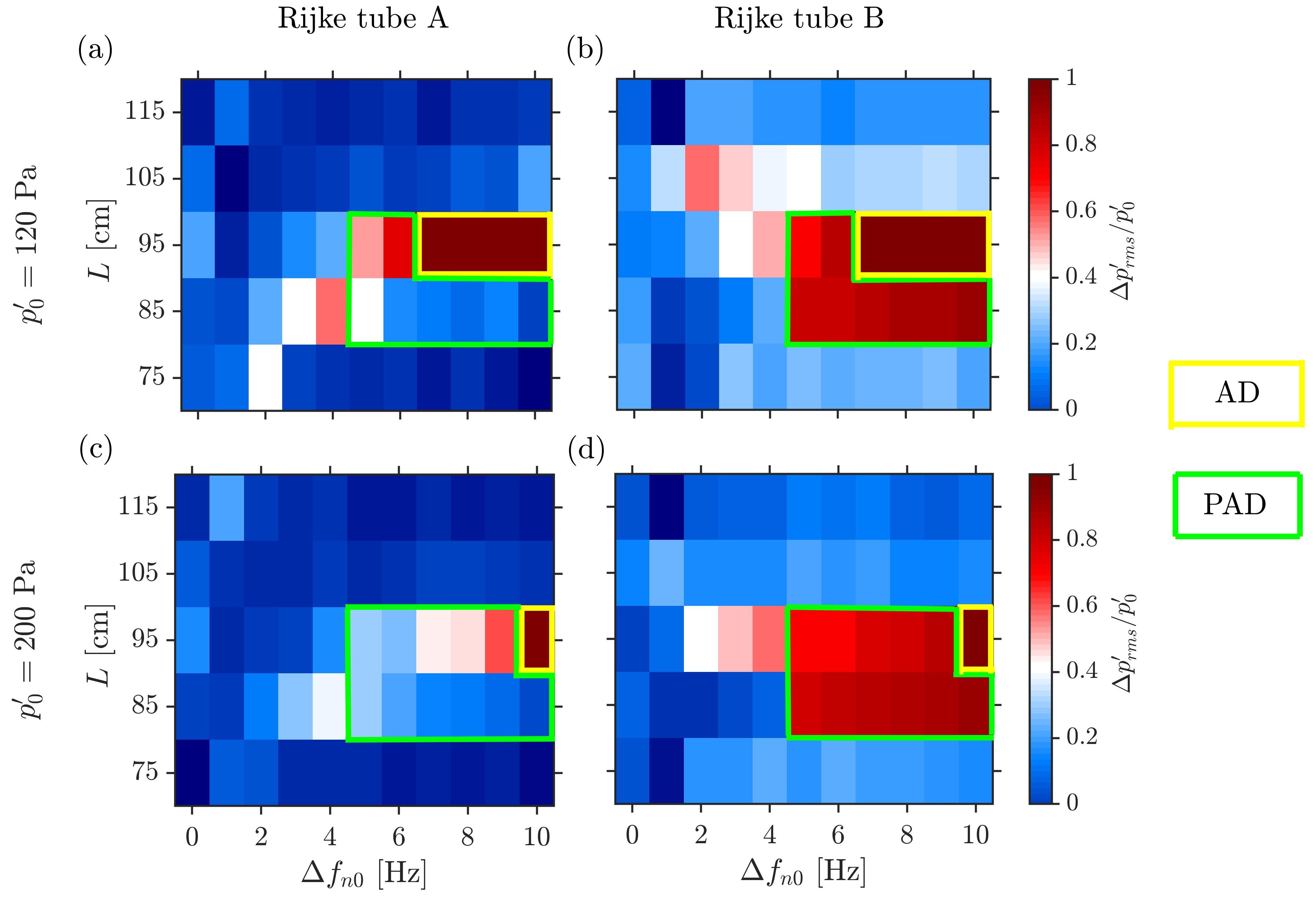}
\caption{\label{l_vs_detuning}Coupled response of two Rijke tubes. Fractional change in the amplitude of LCO for Rijke tube A and B as functions of the frequency detuning ($\Delta f_{n0}=|f_{n0}^A-f_{n0}^B|$) and length of the connecting tube ($L$) for (a,b) $p^\prime_{0}=120$ Pa and (c,d) $p^\prime_{0}=200$ Pa in both the oscillators. The region of AD and PAD are indicated. At other regions in the plots, LCO are observed at reduced amplitude due to coupling. The parametric region exhibiting AD shrinks in size as $p^\prime_{0}$ is increased from $120$ Pa to $200$ Pa.}
\end{figure*}

We also plot the normalized change in the amplitude of LCO due to external forcing, $\Delta p^\prime_{rms}/p^\prime_{0} = (p^\prime_{0}-p^\prime_{rms})/p^\prime_{0}$ in Figs. \ref{single_osc}b,d. Here, $\Delta p^\prime_{rms}/p^\prime_{0} \sim 1$ corresponds to complete suppression of LCO in the Rijke tube due to forcing, whereas $\Delta p^\prime_{rms}/p^\prime_{0}<0$ indicates increase in the amplitude of LCO above the unforced value due to forcing. The green region in Figs. \ref{single_osc}b,d corresponds to an increase in the amplitude of LCO above twice of its unforced value (i.e., $\Delta p^\prime_{rms}/p^\prime_{0}<-1$) due to resonant amplification of the acoustic pressure signal as $f_f$ is very close to $f_{n0}$. $\Delta p^\prime_{rms}/p^\prime_{0}<-1$ implies that the resultant amplitude is more than twice the amplitude of LCO in unforced conditions. The simultaneous occurrence of forced synchronization for $f_f \approx f_{n0}$ along with the resonant amplification of the amplitude of LCO is referred to as synchronance \cite{mondal2019forced}. We further observe a large region with  a reduction in amplitude greater than $80\%$ of $p^\prime_{0}$ only for $f_f < f_{n0}$. A significant decrease in the amplitude of LCO at non-resonant conditions ($f_f \neq f_{n0}$) of forcing is a result of asynchronous quenching \cite{mondal2019forced, guan2019open}. Asynchronous quenching of LCO is achieved through forced synchronization, where the response $p^\prime$ signal oscillates at $f_f$, which can be seen from the coincidence of the boundaries of the Arnold tongue with region corresponding to $\Delta p^\prime_{rms}/p^\prime_{0} \approx 1$ (see Figs. \ref{single_osc}b,d). The asymmetry in the characteristics of the Arnold tongue and asynchronous quenching of LCO arises due to the inherent nonlinearity of the thermoacoustic system \cite{kashinath2018forced}.

\subsection{Response of coupled thermoacoustic oscillators}
\label{Results:2}

We now examine the response of two mutually coupled Rijke tubes when the length of the coupling tube ($L$) and the frequency detuning $\Delta f_{n0}$ $(= f^B_{n0}-f^A_{n0})$ between them are varied independently, where $f^a_{n0}$ and $f^B_{n0}$ are natural frequencies of oscillators A and B in the uncoupled state, respectively. In this study, the diameter of the connecting tube is kept constant at 1 cm for all the experiments (see Section III of Supplemental Material). 

We note that the use of the connecting tube induces an acoustic time delay in the coupling between the two Rijke tubes, as a finite time is required for the acoustic waves in the two oscillators to interact with one another \cite{dange2019oscillation}. A change in the length of the coupling tube changes the acoustic time-delay between the two mutually interacting Rijke tubes. A change in the diameter of the coupling tube changes the coupling strength between the two interacting Rijke tubes. The coupling strength comprises of two parts: time-delay coupling and dissipative coupling. Dissipative coupling is related to the interaction that arises due to mass transfer between the two ducts \cite{bar1985stability}. We observe the presence of both time delay and dissipative coupling in our system.

In Fig. \ref{l_vs_detuning}, we plot the fractional change in the amplitude of LCO in these oscillators, $\Delta p^\prime_{rms}/p^\prime_{0}$, as a function of $L$ and $\Delta f_{n0}$ for two different values of $p^\prime_0=120$ Pa (Figs. \ref{l_vs_detuning}a,b) and $200$ Pa (Figs. \ref{l_vs_detuning}c,d). The colorbar illustrating $\Delta p^\prime_{rms}/p^\prime_{0}$ values range from 0 to 1. During mutual coupling of oscillators, we seldom observe any amplification of LCO.As a result, $\Delta p^\prime_{rms}/p^\prime_{0}$ is never negative. In contrast, during forcing experiments in Fig. \ref{single_osc}b,d, we observe resonant amplification of LCO.

\begin{figure*}
\centering
\includegraphics[width=0.80\textwidth]{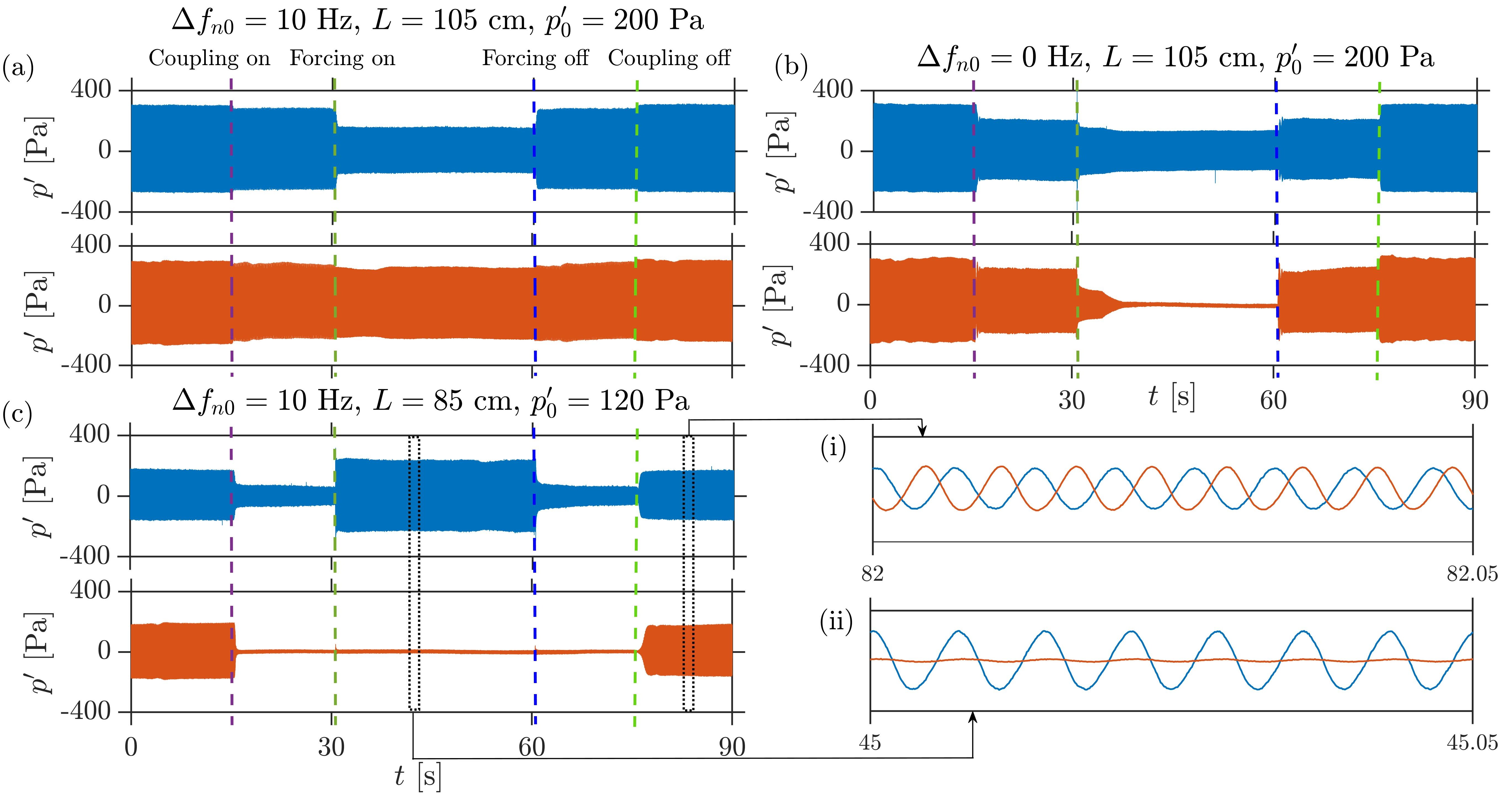}
\caption{\label{long_time_series_all} (a)-(c) Time series of $p^\prime$ in Rijke tubes A (blue) and B (brown) sequentially illustrating the effect of coupling and forcing on the amplitude of LCO in both the Rijke tubes for different coupling and forcing parameters. In (a,c), the Rijke tubes are non-identical and have a frequency difference of 10 Hz. (i,ii) The enlarged portions of the desynchronized LCO and the state of PAD in (c), respectively. The coupling of oscillators always leads reduction in the amplitude of LCO, while forcing of oscillators can have both reduction or amplification effects, depending on the values of coupling and forcing parameters. The common parameters in all plots are: $d=10$ mm, $f_f=140$ Hz, and $A_f=30$ mV.}
\end{figure*}

In Fig. \ref{l_vs_detuning}, for identical oscillators ($\Delta f_{n0}=0$ Hz), we do not observe any perceivable reduction in the amplitude of LCO due to coupling for either values of $p^\prime_0$. However, a significant reduction in the amplitude of LCO is observed when non-identical oscillators are coupled. We observe two states of oscillation quenching, i.e., amplitude death (AD) and partial amplitude death (PAD), in coupled Rijke tubes for a particular range of $L$. For example, when $\Delta f_{n0} = 5-10$ Hz and $L=85$ cm, we witness the presence of PAD in the system (see green box in Fig.  \ref{l_vs_detuning}), while for $\Delta f_{n0}=7-10$ Hz and $L=95$ cm, we notice the occurrence of AD in the system (see yellow box Fig.  \ref{l_vs_detuning}). During the state of PAD, LCO in Rijke tube B (unforced) undergoes suppression and that in Rijke tube A (forced) remains at the reduced amplitude. For the state of AD, we observe greater than $95\%$ decrease in the amplitude of LCO in both the oscillators due to coupling. The range of $\Delta f_{n0}$ for which AD is observed in the system is smaller for $p^\prime_0=200$ Pa (Fig.  \ref{l_vs_detuning}c,d) when compared to that for $p^\prime_0=120$ Pa (Fig.  \ref{l_vs_detuning}a,b). Thus, the occurrence of AD in Rijke tube oscillators has a dependence on the amplitude of uncoupled LCOs. It is quite clear from Fig.  \ref{l_vs_detuning} that the mutual coupling induced through the coupling tube is not capable of inducing AD in identical oscillators, and the occurrence of AD in high amplitude LCO is restricted to a small range of $L$ and requires a finite value of $\Delta f_{n0}$ in the system \cite{dange2019oscillation}.

\subsection{Response of coupled identical thermoacoustic oscillators under asymmetric forcing}
\label{Results:3}

As observed in Fig. \ref{l_vs_detuning}, in the absence of frequency detuning, mutual coupling is ineffective in achieving oscillation quenching (i.e., AD or PAD) in the coupled system. Hence, we asymmetrically force the coupled thermoacoustic oscillators to enhance the quenching of LCOs. Figure \ref{long_time_series_all} shows the representative time series of the acoustic pressure fluctuations in Rijke tube A (in blue) and B (in brown) under the effect of mutual coupling and external forcing when introduced sequentially. Both the oscillators are initially coupled through a single tube and then oscillator A is forced through loudspeakers. The time instant when coupling and forcing are switched on/off are marked as well. In some situations, we notice that coupling and forcing can marginally reduce the amplitude of LCO in the coupled system (Fig. \ref{long_time_series_all}a). While in others, forcing can completely quench LCO in one and reduce the amplitude of LCO in the other oscillator (Fig. \ref{long_time_series_all}b). This situation is akin to the state of PAD. Thus, forcing aids in attaining PAD. On the flip side, in Fig. \ref{long_time_series_all}c, we observe that forcing can also lead to an amplification of the acoustic pressure oscillations, as can be seen from the increased amplitude of these oscillations in Rijke tube A above the uncoupled value. In contrast, the acoustic pressure oscillations remain quenched in Rijke tube B. Thus, the effectiveness of coupling and forcing of thermoacoustic oscillators is restricted to a particular range of coupling and forcing parameters and hence, we focus on the identification of such regions in the subsequent discussion.

\begin{figure*}[t!]
\includegraphics[width=0.80\textwidth]{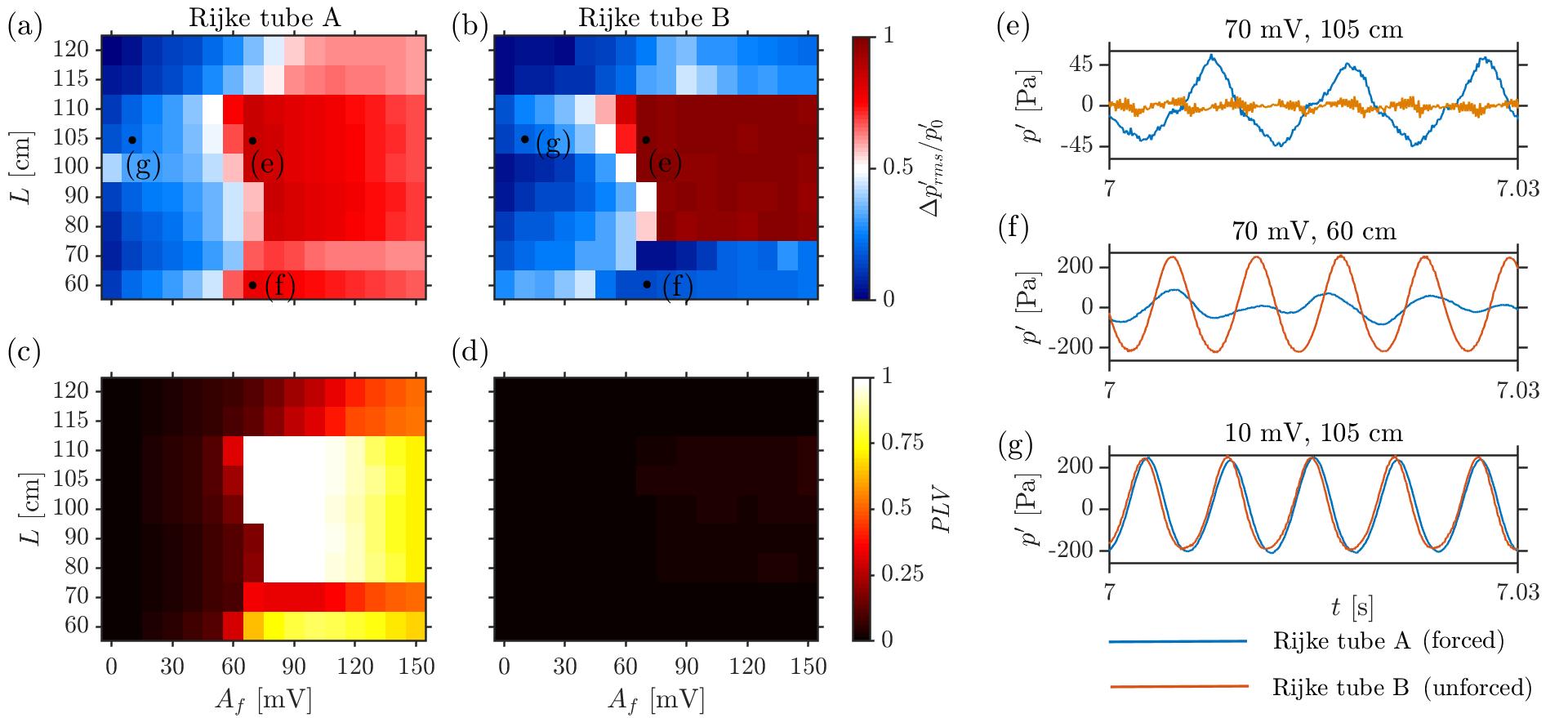}
\caption{\label{l_vs_af_prms_plv} Response of coupled, identical oscillators to asymmetric forcing. (a,b) Amplitude and (c,d) phase response as functions of the forcing amplitude ($A_f$) and length of the coupling tube ($L$). (e-g) Representative time series of $p^\prime$ for the points indicated in (a) depicting difference in the response of Rijke tube A and B. Complementary forcing and coupling enhances the region of $L$ over which the suppression of LCO is observed in both the oscillators. Forcing is ineffective in synchronization of LCO in Rijke tube B (the black region in (d) denotes the complete desynchrony of LCO in Rijke tube B with the forcing signal), while the region of quenching of LCO in Rijke tube A nearly coincides with the region of forced synchronization ($PLV \approx 1$). The common parameters in all the plots are: $p^\prime_{0}=200$ Pa, $f_f=100$ Hz ($f_f/f_{n0} \approx 0.6$), $\Delta f_n = 0$ Hz, and $d=1$ cm.}
\end{figure*}

In Fig. \ref{l_vs_af_prms_plv}, we plot the amplitude (Fig. \ref{l_vs_af_prms_plv}a,b) and the phase (Fig. \ref{l_vs_af_prms_plv}c,d) response of identical Rijke tubes as a function of the forcing amplitude ($A_f$) and the length of coupling tube ($L$). The phase response is determined from the $PLV$ calculated between the external forcing signal and $p^\prime$ for either of the Rijke tubes. Only Rijke tube A is subjected to external forcing. The forcing frequency is chosen as $f_f=100$ Hz ($f_f/f_{n0}\approx 0.6$) for which we observe asynchronous quenching in a single forced Rijke tube oscillator (Fig. \ref{single_osc}d). 

For $A_f=0$, we observe that a change in $L$ does not lead to any suppression of LCO in the oscillators (Fig. \ref{l_vs_af_prms_plv}a,b). With an increase in $A_f$, we observe a gradual decrease in the amplitude of LCO in both the oscillators with the effect on Rijke tube B (unforced) being more pronounced than Rijke tube A (forced). Complete suppression of LCO ($\Delta p^\prime_{rms}/p^\prime_{0} \approx 1$) is observed in a particular range of $L$ in Rijke tube B (Fig. \ref{l_vs_af_prms_plv}b), while a significant decrease in the amplitude of LCO (i.e., $\Delta p^\prime_{rms}/p^\prime_{0} \approx 0.8 $) is observed over a wider range of $L$ in Rijke tube A (Fig. \ref{l_vs_af_prms_plv}a). For the coupled systems without forcing in Fig. \ref{l_vs_detuning}, we did not observe AD or PAD when identical Rijke tubes were coupled for any $L$. However, the same coupled identical oscillators when asymmetrically forced exhibit the state of PAD for a range of $L$ (see point e in Fig. \ref{l_vs_af_prms_plv}a). Thus, we notice that a larger region over which suppression of high amplitude LCO can be achieved in coupled identical thermoacoustic oscillators when the oscillators are coupled in a particular range of L and asymmetric forcing is applied at non-resonant frequencies. We reiterate that we do not observe complete suppression of amplitude of LCO in Rijke tube A. 

Further, we notice that the region of quenching of LCO in Rijke tube A (i.e. $p^\prime_{rms}/p^\prime_{0}>0.8$ in Fig. \ref{l_vs_af_prms_plv}a) nearly coincides with the region of forced synchronization (i.e., $PLV \approx 1$ in Fig. \ref{l_vs_af_prms_plv}c) while the other regions remain desynchronized with the forcing signal. However, the LCO in Rijke tube B always remains desynchronized with the forcing signal, which is observed from the value of $PLV \sim 0$ in Fig. \ref{l_vs_af_prms_plv}d. For the regions where we do not observe suppression of LCO in Rijke tube B, difference between $f_f$ and $f_{n0}$ is too large such that there is no phase locking between oscillations in Rijke tube B and the forcing signal, resulting in low $PLV$. For the regions where we notice complete suppression in Rijke tube B (Fig. \ref{l_vs_af_prms_plv}e), the oscillations are noisy with very low amplitude, leading to low $PLV$.

\begin{figure*}[t!]
\centering
\includegraphics[width=0.75\textwidth]{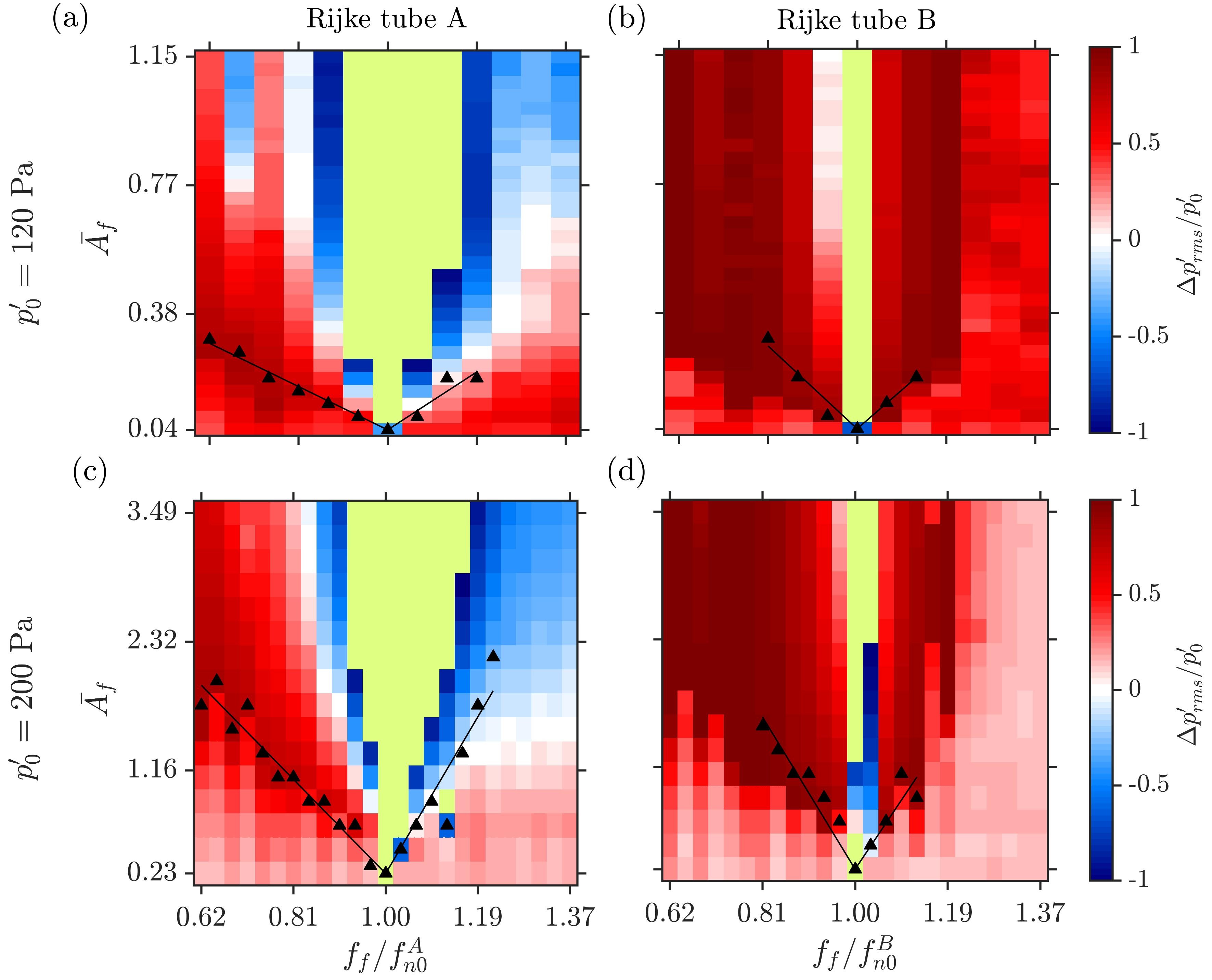}
\caption{\label{asymmetric_zero} Amplitude and phase response (black lines) of Rijke tubes A and B when coupled identical oscillators are subjected to asymmetric forcing for (a,b) $p^\prime_{0}=120$ Pa and (c,d) $p^\prime_{0} = 200$ Pa. The LCO in Rijke tube A is externally forced through the acoustic drivers. The forced synchronization region is wider for Rijke tube A than that observed for Rijke tube B, whereas a much larger magnitude of suppression of LCO is observed in Rijke tube B as compared to that in Rijke tube A. Inside the green region, the oscillations are amplified to values above twice the value of LCO amplitude in the uncoupled-unforced Rijke tubes, such that $\Delta p^\prime_{rms}/p^\prime_{0}$ is observed in the range (-6.47, -1) in (a), (-3.54, -1) in (b), (-6.67, -1) in (c), and (-2.66, -1) in (d). Experimental conditions: $f^A_{n0}=f^B_{n0} \approx 160$ Hz, $L=105$ cm, $d= 1$ cm.}
\end{figure*}

The blue and red time series shown in Fig. \ref{l_vs_af_prms_plv}g depict the acoustic pressure fluctuations in Rijke tubes A and B, respectively and the acoustic pressure time series correspond to data point (g) in Figs. \ref{l_vs_af_prms_plv}a,b. Although these signals appear mutual phase synchronized with each other in Fig. \ref{l_vs_af_prms_plv}g, they are desynchronized with the forcing signal due to a difference in the frequencies of their oscillations (i.e., $f_{n0}= 160$ Hz and $f_f=100$ Hz). Therefore, the PLV distribution quantifying forced synchronization in Fig. \ref{l_vs_af_prms_plv}c and Fig \ref{l_vs_af_prms_plv}d show zero values for data point (g).

Next, we measure the effect of forcing on the Arnold tongue and the amplitude quenching characteristics of the coupled identical Rijke tubes. The length and internal diameter of the coupling tube are fixed at $L=105$ cm and $d=1$ cm, respectively. Only Rijke tube A is forced externally. The natural frequency of both the oscillators is maintained at $f^A_{n0}=f^B_{n0}\approx 160$ Hz. In Fig. \ref{asymmetric_zero}, we depict the fractional change in the amplitude of LCO for each oscillator overlapped with the Arnold tongue on the $\bar{A_f}-f_f$ plane for $p^\prime_{0} = 120$ Pa (Figs. \ref{asymmetric_zero}a,b) and $p^\prime_{0} = 200$ Pa (Figs. \ref{asymmetric_zero}c,d).

\begin{figure*}[t!]
\centering
\includegraphics[width=0.75\textwidth]{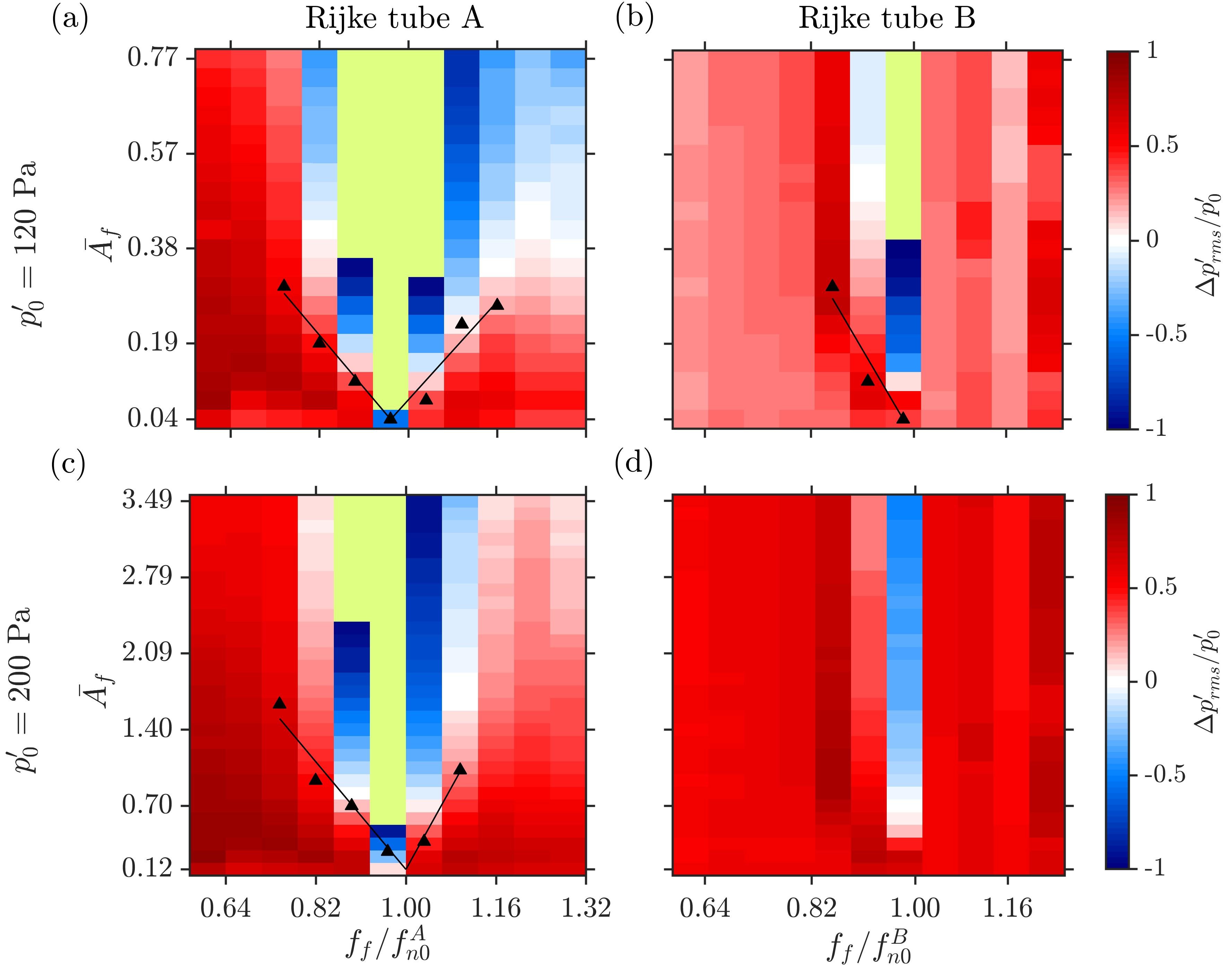}
\caption{\label{10Hz_nonidentical}The amplitude and phase response (black lines) of Rijke tubes A and B when coupled non-identical oscillators are subjected to asymmetric forcing for (a,b) $p^\prime_{0}=120$ Pa and (c,d) $p^\prime_{0} = 200$ Pa. The effect of forcing is less effective in synchronizing and quenching of LCO in Rijke tube B, while it shows regions of forced synchronization and quenching of LCO in Rijke tube A. Inside the region marked in green colour, $\Delta p^\prime_{rms}/p^\prime_{0}$ varies in the range (-6.03, -1) in (a), (-1.47, -1) in (b), and (-3.04, -1) in (c). The common parameters in all the plots are: $\Delta f_{n0}=10$ Hz, $L=105$ cm, $d= 1$ cm.}
\end{figure*}

When the coupled Rijke tubes are asymmetrically forced, we notice forced synchronization at a lower value of $\bar{A_f}$ for Rijke tube A than for Rijke tube B at any value of $f_f$. Therefore, the boundaries of the Arnold tongue are observed to be longer and the forced synchronization region is wider for Rijke tube A (Figs. \ref{asymmetric_zero}a,c) than that for Rijke tube B (Figs. \ref{asymmetric_zero}b,d). Furthermore, we notice that the boundaries of the Arnold tongue for $p^\prime_{0} = 120$ Pa are less steeper than that for $p^\prime_{0} = 200$ Pa. This means that a significantly larger value of forcing amplitude is required to synchronize and quench the large amplitude LCO in the coupled Rijke tube oscillators. As the external forcing is applied to Rijke tube A, the region of resonant amplification is larger than that for Rijke tube B. Moreover, as the forcing amplitude is increased, the oscillator which is subjected to external forcing (i.e., Rijke tube A) synchronizes with the forcing signal first, followed by the forced synchronization of the entire system, i.e., the coupled system of Rijke tubes A \& B. Further, the transition from desynchronized to forced synchronization for Rijke tubes A \& B takes place through a sequence of bifurcations: (i) from desynchronized limit cycle at $f_{n0}^{A,B}$ to two-frequency quasiperiodicity through torus-birth bifurcation ($f_{n0}^{A,B},f_f$); and (ii) from two-frequency quasiperiodicity to synchronized limit cycle ($f_f$) either through saddle-node bifurcation if $f_{n0}^{A,B}$ is close to $f_f$ or through torus-death bifurcation if  $f_{n0}^{A,B}$ is far from $f_f$. These two routes are quite well-known \cite{balanov2008synchronization} and is observed for the identically coupled Rijke tubes under forcing. $f_{n0}$ denotes the natural frequency of the Rijke tubes, and $f_{n}$ denotes the frequency of the acoustic pressure oscillations exhibited by the Rijke tubes after the ducts are coupled and asymmetrically forced.

Rijke tube A (Figs. \ref{asymmetric_zero}a,c) shows a similar trend of quenching of different amplitude LCO as observed for the single oscillator (Figs. \ref{single_osc}b,d). The similarities include an approximate coincidence of maximum amplitude suppression with the synchronization boundary, significant amplitude suppression only for forcing frequencies of $f_f<f_{n0}$, and the presence of resonant amplification region around $f_f/f_{n0} \sim 1$ (shown in green). However, the $\bar{A_f}$ required for forced synchronization at any $f_f$ for the coupled identical oscillators (Figs. \ref{asymmetric_zero}a,c) is higher than that for a single oscillator (Figs. \ref{single_osc}b,d).

In contrast to the response of Rijke tube A, simultaneous effect of coupling and external forcing lead to a much greater suppression of LCO along the boundaries of forced synchronization in Rijke tube B (Figs. \ref{asymmetric_zero}b,d). Suppression of LCO in Rijke tube B is observed for both $f_f<f_{n0}$ and $f_f>f_{n0}$, unlike Rijke tube A where we notice a reduction only for $f_f<f_{n0}$. Further, the range of $f_f$ and $\bar{A_f}$ over which the suppression of LCO occurs in Rijke tube B is larger than that observed for Rijke tube A. Thus, we reassert that asymmetrically forced coupled system exhibits suppression of higher amplitude LCO for a larger range of forcing parameters than that observed when the oscillators are forced or coupled individually.

\subsection{Forced response of coupled non-identical limit cycle oscillators}
\label{Results:4}

Now, we study the response of coupled non-identical Rijke tubes under asymmetric forcing. Only Rijke tube A is externally forced. In Fig. \ref{10Hz_nonidentical}, we depict the Arnold tongue and the fractional change in the amplitude of LCO for $p^\prime_{0}=120$ Pa (Figs. \ref{10Hz_nonidentical}a,b) and $p^\prime_{0}=200$ Pa (Figs. \ref{10Hz_nonidentical}c,d). The oscillators are coupled through a tube of $L=105$ cm and $d=1$ cm. A frequency detuning between the uncoupled oscillators is fixed at 10 Hz (i.e., $\Delta f_{n0} = f^B_{n0}-f^A_{n0}=10$ Hz). We notice that the effect of forcing is quite insignificant for synchronization of LCO in Rijke tube B. We observe a small range of $f_f$ (i.e., $f_f<f^B_{n0}$) over which Rijke tube B is synchronized to forcing for $p^\prime_{0} = 120$ Pa (Fig. \ref{10Hz_nonidentical}b). The region of forced synchronization is completely absent when $p^\prime_0$ is increased to 200 Pa (Fig. \ref{10Hz_nonidentical}d). In contrast, as the external forcing is applied directly to Rijke tube A, it is easily synchronized with the forcing, as seen from the longer boundaries of the Arnold tongue for both $p^\prime_{0} = 120$ and 200 Pa in Figs. \ref{10Hz_nonidentical}a,c, respectively. 

Finally, the mechanism through which forced sychronization is attained for $p^\prime_{0} = 120$ Pa in the non-identical system is quite different from that of the coupled system. As the forcing amplitude is increased, the oscillator which is subjected to external forcing (i.e., Rijke tube A) synchronizes with the forcing signal first, followed by the forced synchronization of the entire system, i.e., the coupled system of Rijke tubes A \& B. The entire system undergoes forced synchronization only in $f_f/f_{n0}^{A,B}< 1$ range. The transition from desynchronized to forced synchronization for Rijke tube A takes place through the following sequence of bifurcations: (i) from desynchronized two-frequency quasiperiodicity ($f_{n0}^A,f_{n0}^B$) to three-frequency quasiperiodicity ($f_{n0}^A,f_{n0}^B,f_f$) as $A_f$ is increased; and (ii) from three-frequency quasiperiodicity to synchronized limit cycle ($f_f$). This route was reported for coupled Van der Pol oscillators under asymmetric forcing \cite{anishchenko2008bifurcational,anishchenko2009phase}. However, Rijke tube B goes from two-frequency quasiperiodicity ($f_{n0}^A,f_{n0}^B$) to synchronized limit cycle ($f_f$) when $f_f/f_{n0}<1$. For $p^\prime_{0} = 200$ Pa, the acoustic pressure fluctuations in Rijke tube B remain desynchronized with the forcing signal throughout the $\bar{A_f}-f_f$ parameter plane.

The amplitude response of Rijke tube A and B shows that the significant suppression of LCO can still be achieved at non-resonant conditions of forcing. The region of resonant amplification observed around $f_f/f_{n0} \sim 1$ is very small for Rijke tube B as compared to Rijke tube A. Furthermore, the comparison of the forced response of the coupled oscillators with $p^\prime_{0}=120$ Pa (Figs. \ref{10Hz_nonidentical}a,b) and $p^\prime_{0}=200$ Pa (Figs. \ref{10Hz_nonidentical}c,d) shows that  for the higher amplitude LCO, we need significantly larger values of $\bar{A_f}$ for synchronization and quenching of oscillations in both the Rijke tubes. This is similar to the observations made for identical oscillators in Fig. \ref{asymmetric_zero}. Note that the ordinate in Figs. \ref{10Hz_nonidentical}c,d is much larger than that in Figs. \ref{10Hz_nonidentical}a,b. Thus, we re-emphasize that the effect of forcing is more significant in suppressing LCO in both coupled thermoacoustic oscillators if their natural frequencies are nearly the same as compared to that seen in the case of non-identical oscillators.

\section{Mathematical model}
\label{S4:model}

In this section, we will discuss a reduced-order model developed for the system of coupled horizontal Rijke tubes subjected to asymmetric forcing. The model of the uncoupled oscillator is based on \citet{balasubramanian2008thermoacoustic}. We neglect the effects of mean flow and mean temperature gradient in the duct. The temporal evolution of a single Rijke tube is described by the following set of ODEs
\begin{equation}  \label{eq:10}
    \dfrac{d \eta_j}{dt} = \Dot{\eta_{j}},
\end{equation}
\begin{equation}  \label{eq:11}
\begin{split}
    \dfrac{d \Dot{\eta_{j}}}{dt} &  + 2 \xi_j \omega_j \Dot{\eta_{j}} + k_{j}^2 \eta_{j} \\ & = -j \pi K \left[\sqrt{\left| \dfrac{1}{3} + u_{f}^\prime(t-\tau) \right|} -\sqrt{\dfrac{1}{3}}\right]\hspace{2 pt}\text{sin}(j\pi x_f),
\end{split}
\end{equation} where $k_j = j\pi$ refers to the non-dimensional wave number and $\omega_j = j \pi$ refers to the non-dimensional angular frequency of the $j^{\text{th}}$ mode. Other parameters are the non-dimensional heater power $K$, and non-dimensional velocity, $u^\prime_{f}$ at the  non-dimensional heater location, $xf$. The thermal inertia of the heat transfer in the medium is captured by a parameter time lag $\tau$.The coefficient $\xi_j$ appearing in the second term of Eq. (\ref{eq:11}) represents the frequency-dependent damping \cite{matveev2003thermoacoustic}, and is given by the following ansatz \cite{sterling1991nonlinear}:

\begin{equation} \label{eq:12}
    \xi_j = \dfrac{c_1 \dfrac{\omega_j}{\omega_1}+c_2\sqrt{\dfrac{\omega_1}{\omega_j}}}{2\pi}
\end{equation}
Here, $c_1$ and $c_2$ are the damping coefficients which determine the amount of damping. We choose the values $c_1=0.1$ and $c_2=0.06$ based on \cite{sterling1991nonlinear} for all simulations. We also set $x_f=0.25$ ($=L/4$), as it is the most favorable location for the onset of thermoacoustic instability \cite{matveev2003thermoacoustic}.
The non-dimensional velocity $u^\prime$ and non-dimensional pressure $p^\prime$ fluctuations in the model are written in terms of the Galerkin modes:

\begin{equation} \label{eq:8}
    p^\prime(x,t) =  \sum_{j=1}^{N} -\dfrac{\gamma M}{j \pi}\dot{\eta_{j}}(t)\hspace{2 pt}\text{sin}(j \pi x),
\end{equation}

\begin{equation} \label{eq:9}
    u^\prime(x,t) =  \sum_{j=1}^{N} \eta_{j}(t)\hspace{2 pt}\text{cos}(j \pi x).
\end{equation}

For Rijke tubes, \citet{matveev2003thermoacoustic} and \citet{sayadi2013thermoacoustic} have shown that the first mode is the most unstable of all the other modes. Consequently, we consider only the first mode ($N=1$) in our numerical analysis, to keep the model simple.

\begin{figure*}[t!]
\centering
\includegraphics[width=0.75\textwidth]{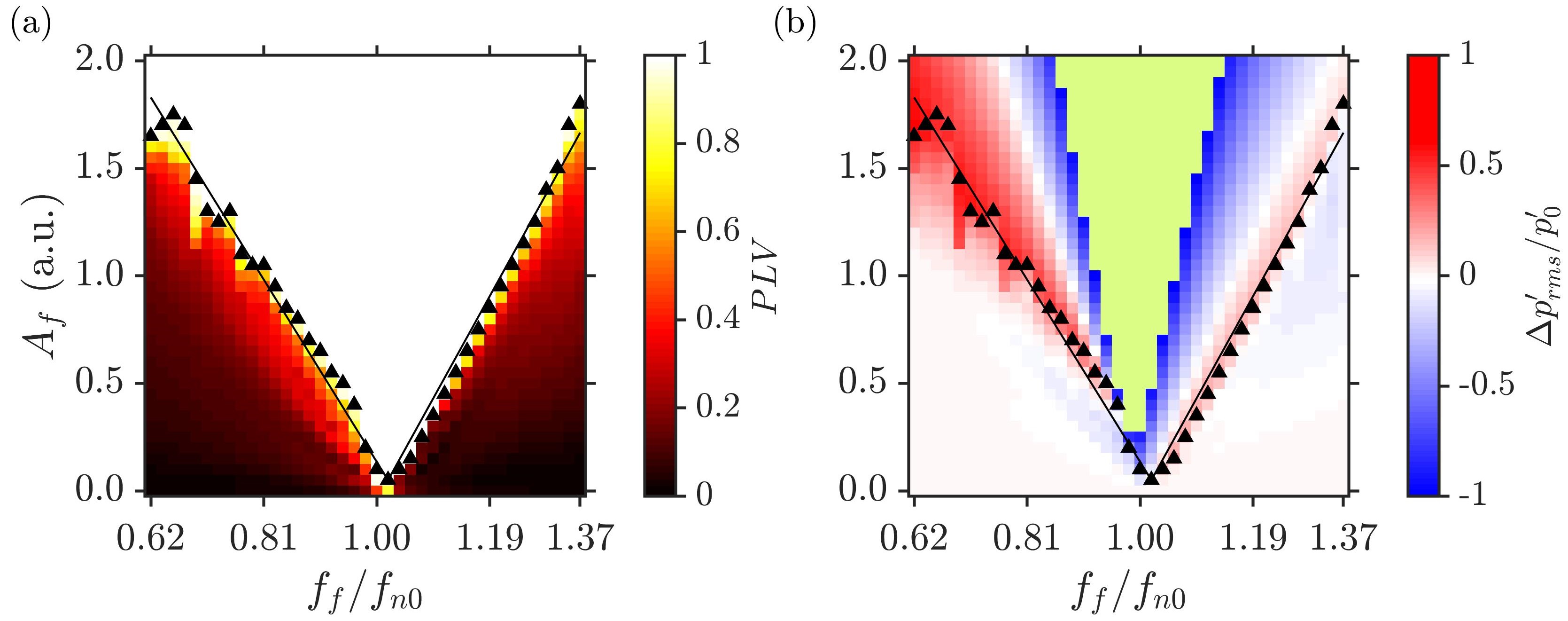}
\caption{\label{single_osc_model}(a) The phase response and (b) the amplitude response obtained from the model of a single Rijke tube under external forcing. The synchronization boundaries are obtained through a least-square-fit of points where $PLV=0.98$. The Arnold tongue and extent of asynchronous quenching from the model show a qualitative match with the experimental results in Fig. \ref{single_osc}. In (b), the green region around $f_f/f_{n0} = 1$ indicates doubling of LCO amplitude from its unforced value, where $\Delta p^\prime_{rms}/p^\prime_{0}$ varies in the range (-6.35, -1) in (b).}
\end{figure*} 

Let superscripts ``$a$" and ``$b$" denote Rijke tube A and B, respectively. We assume that the two Rijke tubes are coupled through time-delay and dissipative couplings. The governing equations for coupled non-identical Rijke tubes with asymmetric sinusoidal forcing can then be written as:

\begin{equation} \label{eq:13}
    \dfrac{d \eta_j^{a,b}}{dt} = \Dot{\eta_{j}}^{a,b},
\end{equation}

\begin{multline} \label{eq:14}
\dfrac{d \Dot{\eta_{j}}^{a,b}}{dt} + 2 \xi_j \left(\dfrac{\omega_j}{r^{a,b}}\right) \Dot{\eta_{j}}^{a,b} + \left(\dfrac{k_{j}}{r^{a,b}}\right)^2 \eta_{j}^{a,b} \\ = -\dfrac{j \pi}{{r^{a,b}}^2} K \left[\sqrt{\left| \dfrac{1}{3} + u_{f}^{\prime a,b} (t-\tau) \right|} -\sqrt{\dfrac{1}{3}}\right]\hspace{2 pt}\text{sin}\left(\dfrac{j\pi x_f}{r^{a,b}}\right) \\ + \underbrace{K_d(\dot{\eta_j}^b - \dot{\eta_j}^a)}_\text{Dissipative coupling} + \underbrace{K_\tau (\dot{\eta_j}^b(t-\tau_{tube})-\dot{\eta_j}^a(t))}_\text{Time-delay coupling} \\ + \underbrace{[A_f \hspace{2 pt} \text{sin}(2 \pi f_f t)]^a}_\text{Forcing term},
\end{multline}

where, $r^{a,b}$ is defined as the ratio of the length of the duct to a reference length, $L_{a,b}/L_{ref}$. We consider $L_a$ to be the reference length. For identical oscillators, $r^a=r^b = 1$. For non-identical oscillators, $r_a = 1$ and $r_b = L_b/L_a = \omega_a/\omega_b$. The detailed derivation of Eq. (\ref{eq:14}) is provided in Section IV of the Supplemental Material.

The second and third terms on the right-hand side of Eq. (\ref{eq:14}) are the dissipative and time-delay coupling terms, respectively. Dissipative coupling encapsulates the interaction that arises from the mass transfer between the two ducts \cite{bar1985stability}. Time-delay coupling quantifies the time taken by acoustic waves to propagate through the coupling tube connecting the two Rijke tubes \cite{biwa2015amplitude, thomas2018effect}. Thus, $\tau_{tube}$ denotes the time-delay in the response induced by one Rijke tube on the other, and is proportional to the length of the coupling tube, i.e., $\tau_{tube} \propto L/c$, where $c$ is the speed of sound inside the coupling tube. The fourth term is the sinusoidal forcing term with amplitude $A_f$ and frequency $f_f$. The external forcing is applied only to Rijke tube A. 

The effect of the source term is characterized by the time-delay $\tau$, which captures the thermal inertia of the heat transfer in the medium. The ODEs given in Eqs. (\ref{eq:13}) and (\ref{eq:14}) are solved numerically using \texttt{dde23}, an inbuilt function for solving delay differential equations in MATLAB \cite{shampine2001solving}, and $p'$ is calculated using Eq. (\ref{eq:8}).  The parameters which are kept constant during the simulations are indicated in Table \ref{tab:table-name}. These parameters have been chosen such that the both the oscillators exhibit limit cycle oscillations at a parametric location away from the bistable regime and also to quantitatively match the experimental results.

\begin{table}[t!]
\centering
\begin{tabular}{cc|cc|cc}
\hline
\hline
Parameter & Value & Parameter & Value & Parameter & Value \\
\hline
$N$ & 1 & $\tau$ & 0.25 & $K$ & 5 \\
$\gamma$ & 1.4 & $M$ & 0.005 & $x_f$ & 0.25 \\
$c_1$ & 0.1 & $c_2$ & 0.06 & $\omega_f$ & $j\pi$ \\
\hline\hline
\end{tabular}
\caption{\label{tab:table-name}Model parameters kept constant throughout the numerical analysis of the reduced-order model.}
\end{table}

\subsection{Model results for a single oscillator and mutually coupled oscillators}%Forced response of an isolated Rijke tube: Forced synchronization and amplitude quenching}
\label{model_single}

\begin{figure*}[t!]
\centering
\includegraphics[width=0.80\textwidth]{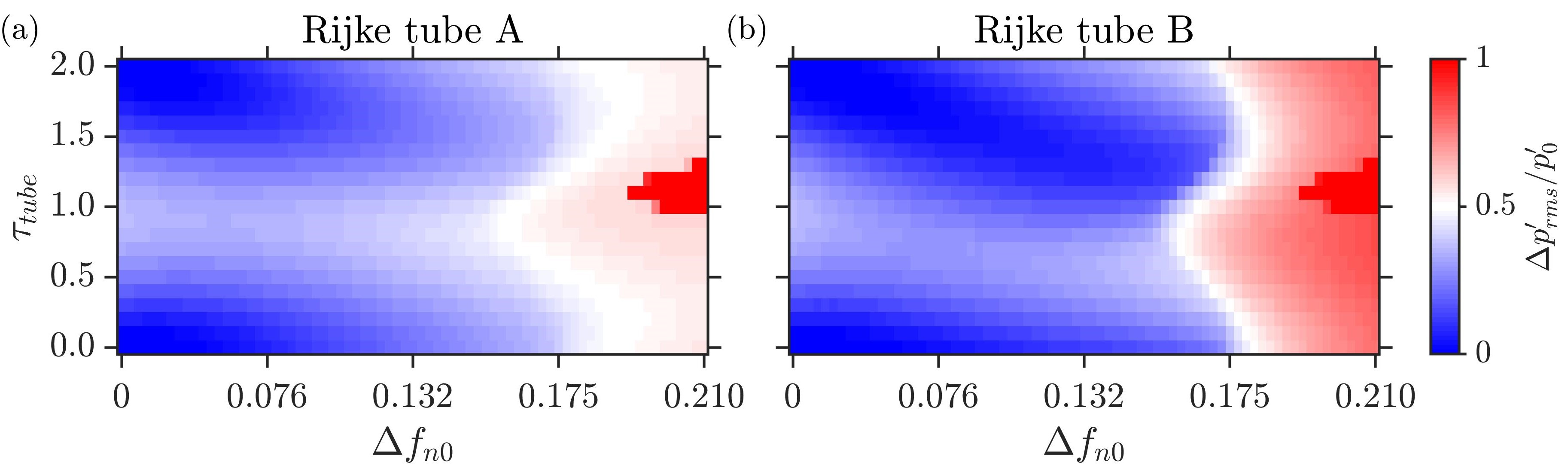}
\caption{\label{det_vs_tau}Two-parameter bifurcation plots between frequency detuning between the two Rijke tube model oscillators and $\tau_{tube}$. $K_d$ and $K_{\tau}$ are kept constant at $1.0$ and $0.2$, respectively. The dark red region in both the plots represent regions of AD observed in the model oscillators. At other places, LCO are observed at reduced amplitude due to coupling. }
\end{figure*}

\begin{figure*}[t!]
\centering
\includegraphics[width=0.80\textwidth]{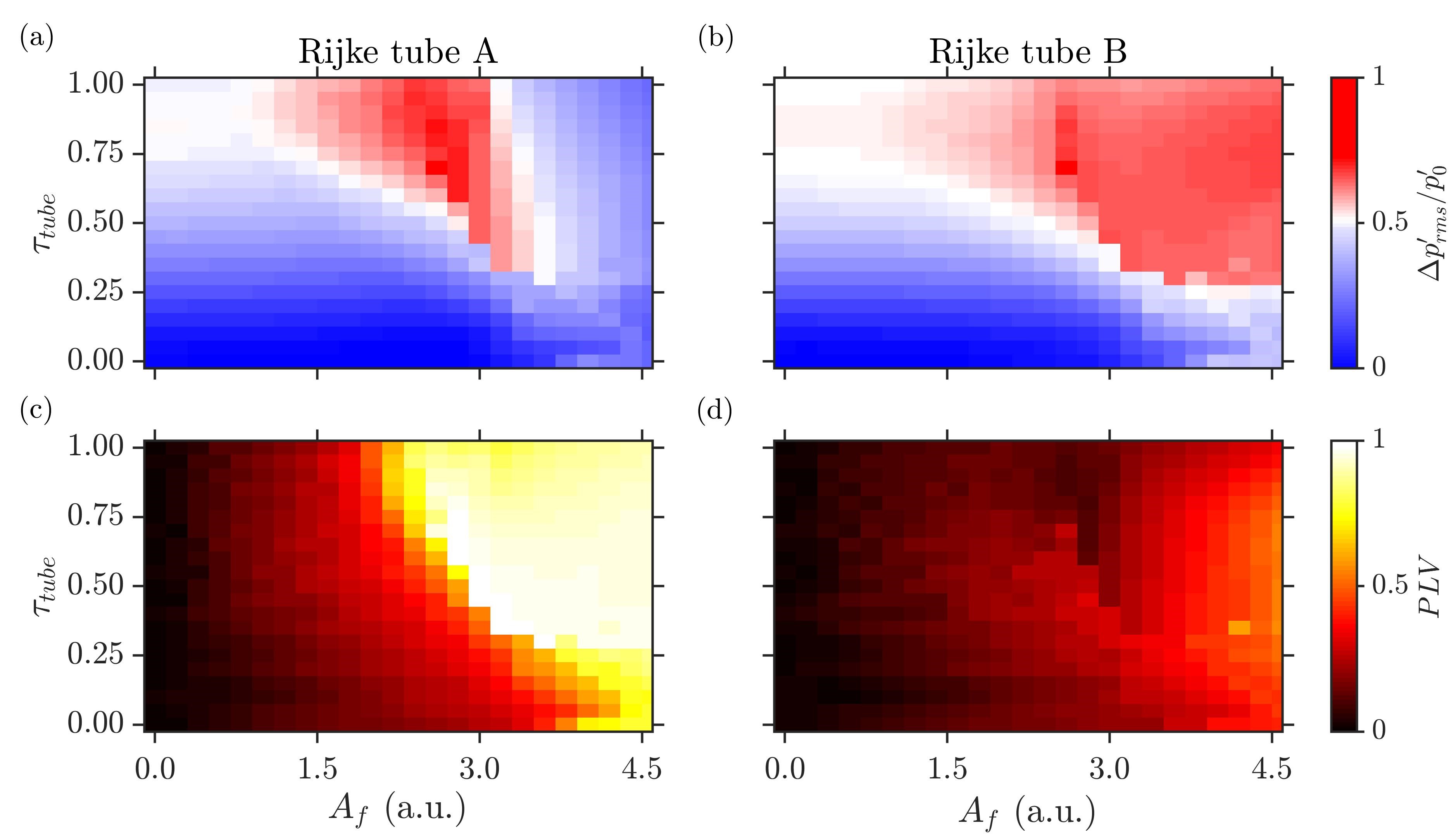}
\caption{\label{Af_vs_kt} (a,b) Amplitude and (c,d) phase response of coupled, identical oscillators under asymmetric forcing for different values of $\tau_{tube}$ and $A_f$. Amplitude response is indicated by the fractional change in the amplitude of LCO ($\Delta p^\prime_{rms}/p^\prime_{0}$) while phase response is quantified by the $PLV$ between $p^\prime$ and forcing signal. For $\tau_{tube}>0.4$ and $A_f>1.5$, large suppression in the amplitude of LCO is observed in the two Rijke tubes.}
\end{figure*}

We begin by analyzing the forced response of a single Rijke tube by adding the forcing term $A_f\sin(2 \pi f_f t)$ to the right hand side Eq. (\ref{eq:11}). Considering only the first mode, we get the following set of ODEs is:
\begin{equation}  \label{eq:15}
    \dfrac{d \eta_1}{dt} = \Dot{\eta_{1}},
\end{equation}
\begin{equation}  \label{eq:16}
\begin{split}
\dfrac{d \Dot{\eta_{1}}}{dt} & + 2 \xi_1 \omega_1 \Dot{\eta_{1}} + k_{1}^2 \eta_{1} = - \pi K \left[\sqrt{\left| \dfrac{1}{3} + u_{f}^\prime(t-\tau) \right|} -\sqrt{\dfrac{1}{3}}\right] \\ & \times \hspace{2 pt}\text{sin}(\pi x_f)  + \underbrace{A_f \hspace{2 pt} \text{sin}(2 \pi f_f t)}_\text{Forcing term}.
\end{split}
\end{equation}

Figure \ref{single_osc_model} shows the two-parameter bifurcation plot on the $A_f-f_f$ plane illustrating the phase and amplitude dynamics of the limit cycle oscillations in a single Rijke tube, obtained from the mathematical model. We observe the existence of Arnold tongue in Fig. \ref{single_osc_model}a along with the region of asynchronous quenching ($\Delta p^\prime_{rms}/p^\prime_{0}>0.5$) in Fig. \ref{single_osc_model}b. We observe the qualitative match between experimental results shown in Fig. \ref{single_osc} and those obtained from the model in Fig. \ref{single_osc_model}. However, there are a number of differences. First, the region of asynchronous quenching observed in the model is smaller than that observed in the experiments. Second, the resonant amplification region is spread over a larger extent of parameter values in the model. Finally, the Arnold tongue obtained from the model is symmetrical as opposed to left-skewed in the experiments. The skewness of the Arnold tongue increases as the model heater power $K$ is increased to higher-values, indicating the nonlinear behavior of the overall systems. Figure \ref{single_osc_model} has been shown for a model heater power $K$ for which we get the best match with the experimental results in Fig. \ref{single_osc}.

In Fig. \ref{l_vs_detuning}, we showed that the state of AD was attained in coupled Rijke tubes only when there was a frequency detuning present in the two oscillators, in addition to the dissipative and time-delay coupling due to the connecting tube. In Fig. \ref{det_vs_tau}, we show the effect of coupling and frequency detuning on the occurrence of the state of AD in the model. We fixed the value of $K_d=1.0$ and $K_\tau=0.2$, and varied $\tau_{tube}$ and frequency detuning $\Delta f_{n0}$. For identical oscillators ($\Delta f_{n0}=0$), we did not observe any appreciable reduction in the amplitude of LCO. However, for finite value of detuning, we observe the state of AD in the model, consistent with our experimental observations in Fig. \ref{l_vs_detuning}. Thus, the effect of change of the length of coupling tube is captured quite well by the time-delay $\tau_{tube}$.

The results summarized in Figs. \ref{single_osc_model} and \ref{det_vs_tau} validate our model against experimental observations for a single and coupled Rijke tubes reported in the present study (Figs. \ref{single_osc} and \ref{l_vs_detuning}), as well as those made in past studies \cite{thomas2018effect, hyodo2018stabilization, dange2019oscillation}. We now turn our attention towards modeling the effect of asymmetric forcing on the coupled Rijke tubes. 

\subsection{Model results for coupled behavior of thermoacoustic oscillators under asymmetric forcing}
\label{model_coupled_forced}

\begin{figure*}[t!]
\centering
\includegraphics[width=0.75\textwidth]{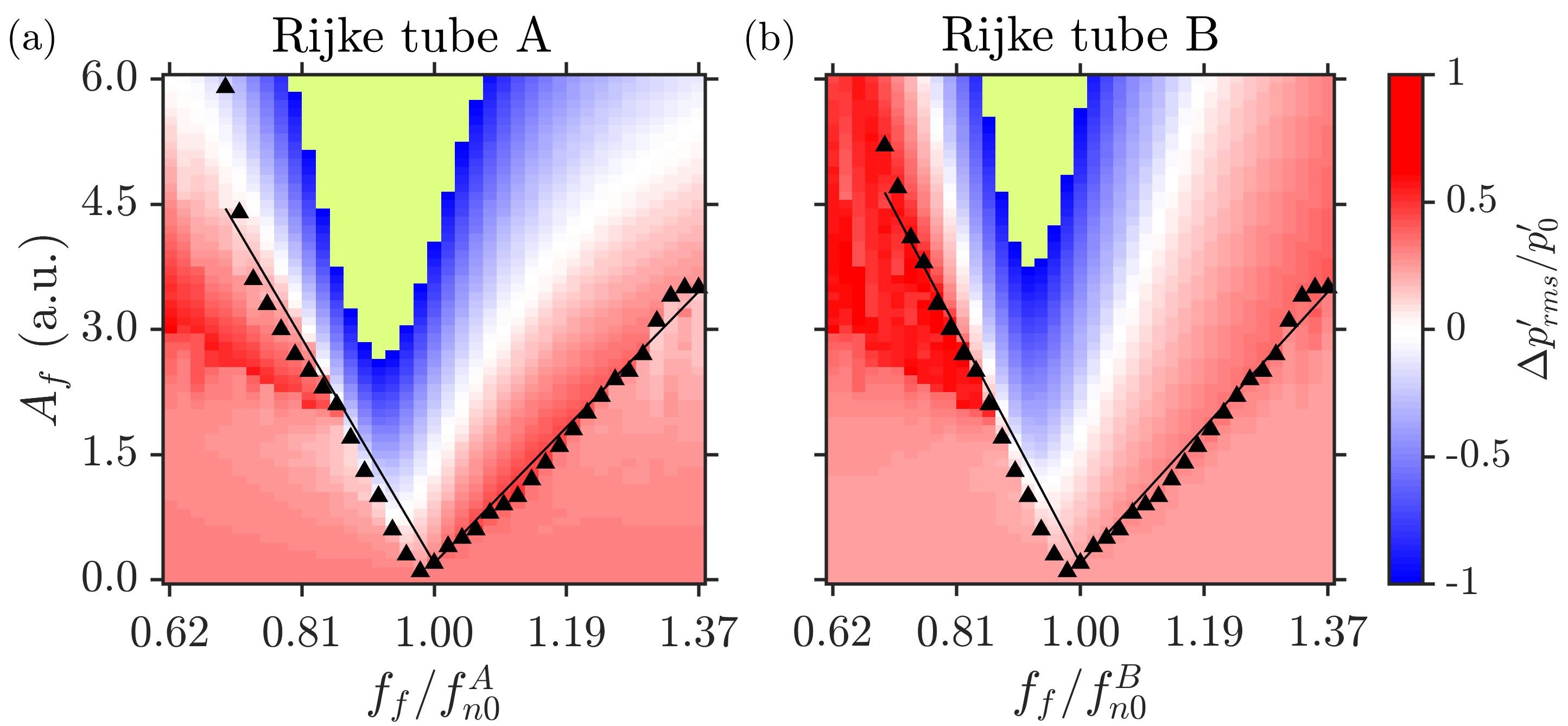}
\caption{\label{coupled_identical}The amplitude response and overlapped Arnold tongue for identical, coupled Rijke tubes from the numerical model. External forcing is applied to Rijke tube A. The synchronization boundaries are obtained through least-square-fit of points where $PLV=0.98$. For the green region around $f_f/f_{n0} = 1$, values of $\Delta p^\prime_{rms}/p^\prime_{0}$ vary in the range (-3.20, -1) in (a), and (-2.08, -1) in (b). The coupling parameters are: $K_d = 1.0$, $K_{\tau} = 0.2$ and $\tau_{tube} = 0.4$.}
\end{figure*}

In Fig. \ref{Af_vs_kt}, we plot the amplitude and the phase response of the coupled identical Rijke tubes under asymmetric forcing, i.e., only Rijke tube A is subjected to external forcing according to Eq. (\ref{eq:14}). The effect of coupling is parameterized by $\tau_{tube}$, while that of external forcing by $A_f$. The forcing frequency is fixed at $f_f/f_{n0} = 0.6$. $K_d$, $K_{\tau}$ \& $K$ values are fixed at $1.0$, $0.2$ and $5.0$, respectively. Note that the heater power $K=5.0$ leads to high amplitude LCO in the two Rijke tubes ($p^{{\prime}{A,B}}_0 = 648$ Pa).

For $A_f = 0$, we notice that the change in $\tau_{tube}$ leads to very low suppression in either of the oscillators ($\Delta p^\prime_{rms}/p^\prime_{0}<0.5$). This is again a reflection of the fact that the region of AD is quite limited when identical oscillators are coupled. Our aim is to illustrate that external forcing can lead to significant quenching of LCO in the identical oscillators. When $A_f$ is increased, we observe a decrease in the amplitude of LCO in both the oscillators after a critical value of $A_f$. The effect of forcing is more pronounced in Rijke tube B as compared to that in Rijke tube A, which is evident from the larger ``red" region in Fig. \ref{Af_vs_kt}b compared to that in Fig. \ref{Af_vs_kt}a. %We observe that for $K_{\tau}>0.20$ values, a large suppression ($\Delta p^\prime_{rms}/p^\prime_{0} \sim 0.70$) of the LCO is observed in Rijke tube B. Based on this, $K_{\tau}$ is fixed at $0.20$ for subsequent simulations.

The results obtained from the model (Fig. \ref{Af_vs_kt}) can be compared with the experimental results shown in Fig. \ref{l_vs_af_prms_plv}. The ratio of forcing frequency to natural frequency is kept same for ease of comparison. Although the magnitude of amplitude suppression is not captured quantitatively by the model, the qualitative trends such as the extent of amplitude suppression in the two model oscillators along with the forced synchronization characteristics are consistent between the experimental and numerical results. Compared to the pressure oscillations in Rijke tube B exhibiting complete suppression in the experimental study (for e.g., point (e) in Fig. \ref{l_vs_af_prms_plv}a), we observe around $70 \%$ suppression in the amplitude of LCO in Rijke tube B. 

Similar to experimental results, we also observe that the LCO in Rijke tube B remains desynchronized with the forcing signal through a large extent of the parameter plane. The region of quenching of LCO in Rijke tube A ($\Delta p^\prime_{rms}/p^\prime_{0} \sim 1$) nearly coincides with the onset of forced synchronization ($PLV \sim 1$). The results shown in Fig. \ref{Af_vs_kt} are in qualitative agreement with the experimental results indicated in Figs. \ref{l_vs_af_prms_plv}a-d.

In Fig. \ref{coupled_identical}, we depict the fractional change in the amplitude of LCO for each oscillator overlapped with the Arnold tongue on the $A_f - f_f$ plane, obtained from the numerical simulations for the case where identical oscillators are asymmetrically forced. The values of the coupling constants are fixed at $K_d = 1.0$, $K_{\tau} = 0.20$ and $\tau_{tube} = 0.4$. The normalized natural frequency of both the oscillators is $0.5$, and the normalized heater power are maintained at $K=5.0$. 

We see that the amplitude dynamics observed during the experiments, as depicted in Fig. \ref{asymmetric_zero}, is well captured by the numerical model in Fig. \ref{coupled_identical}. From the numerical results, we observe larger magnitude of suppression of LCO in Rijke tube B, as compared to that in Rijke tube A. Also, the high magnitude of suppression ($\Delta p^\prime_{rms}/p^\prime_{0}>0.5$) in Rijke tube B nearly coincides with the boundary of Arnold tongue in Rijke tube A. The resonant synchronization region in Rijke tube B is smaller in size compared to that in Rijke tube A. The model does not adequately capture the phase dynamics as observed in the experiments. Although the Arnold tongue is similar for both the Rijke tubes, the $PLV$ values observed prior to the onset of forced synchronization is lower in Rijke tube B than that in Rijke tube A. Also, the resonant amplification region (shown in green) is skewed towards the left side of $f_f/f_{n0} = 1.00$ value. This is different from the experimental results (Fig. \ref{asymmetric_zero}), where the resonant amplification region was almost symmetrically located around the $f_f/f_{n0} = 1.00$ value.

\begin{figure*}[t!]
\centering
\includegraphics[width=0.75\textwidth]{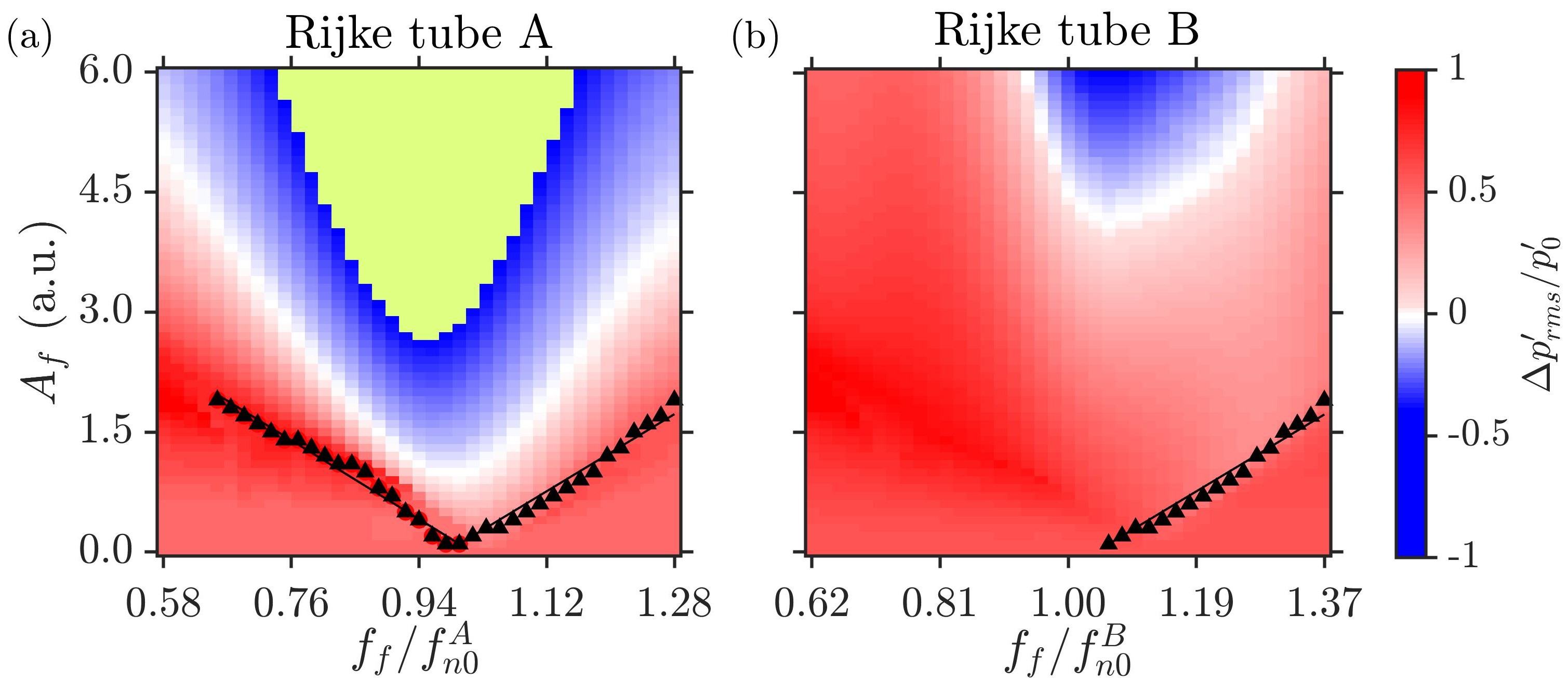}
\caption{\label{coupled_non_identical}The amplitude response and overlapped Arnold tongue for non-identical, coupled Rijke tubes from the numerical model. External forcing is applied to Rijke tube A. The synchronization boundaries in are obtained through least-square-fit of points where $PLV=0.98$. Inside the green region in (a), $\Delta p^\prime_{rms}/p^\prime_{0}$ varies in (-2.66, -1) range. The coupling parameters are: $K_d = 1.0$, $K_{\tau} = 0.2$ and $\tau_{tube} = 0.4$.}
\end{figure*}

Finally, we use our model to characterize the response of asymmetric forcing on the Arnold tongue and amplitude characteristics of coupled non-identical Rijke tubes. The coupling constants and non-dimensional heater power values are kept same as those used to get the results shown in Fig. \ref{coupled_identical},i.e., $K_d = 1.0$, $K_{\tau} = 0.2$, $\tau_{tube} = 0.4$, and $K = 5$. This ensures the consistency of initial and boundary conditions, as was ensured in the experiments. In  \S \ref{Results:4}, to study the forced response of coupled non-identical Rijke tube oscillators, the natural frequencies of the LCO in Rijke tubes A and B were maintained at $150$ Hz and $160$ Hz, respectively, which equals a frequency ratio of $160/150 = 1.067$. Consistent with the experiments, the value of $r$ is kept as $1.066$ to analyze the reduced-order model given in Eqs. (\ref{eq:13}) and (\ref{eq:14}).

We observe that the effect of forcing is quite insignificant for forced synchronization of LCO in Rijke tube B, consistent with the experimental results (Fig. \ref{10Hz_nonidentical}). A significant suppression of the amplitude of LCO is observed at non-resonant conditions of forcing in Rijke tube A. The region of resonant amplification is absent for Rijke tube B, although we do observe amplification of the amplitude of LCO at higher values of $A_f$ in Fig. \ref{coupled_non_identical}b. Some differences are present between the experimental (Fig. \ref{10Hz_nonidentical}) and numerical results (Fig. \ref{coupled_non_identical}). First, the forced synchronization of Rijke tube B is observed for $f_f/f_{n0}^B < 1$ values during experiments, whereas, the pressure oscillations in Rijke tube B model oscillator undergo forced synchronization for $f_f/f_{n0}^B > 1$ values. Second, the Arnold tongue is spread over a larger extent of parameter values for a coupled non-identical oscillators system as opposed to that in a coupled identical oscillator system (Fig. \ref{coupled_identical}). During experiments, an opposite behavior was observed in Fig. \ref{10Hz_nonidentical}c, where the size of the Arnold tongue of Rijke tube A for a coupled non-identical system is less when compared to that in the coupled identical system (Fig. \ref{asymmetric_zero}c). The qualitative differences between the results can be attributed to the various assumptions used while deriving the reduced-order model, such as absence of mean flow, zero temperature gradient, non-heat conducting gas, etc. A more detailed study is needed to examine the effects of such assumptions on the numerical results. 

\section{Conclusions}
\label{Conclusions}

In this proof-of-concept study, we investigated the phase and the amplitude dynamics of coupled thermoacoustic oscillators under asymmetric forcing, and presented a model which satisfactorily captured the experimental results qualitatively. In particular, we discuss the viability of simultaneous coupling and forcing as a method for controlling thermoacoustic instability in a system of multiple combustors.

The forced response of limit cycle oscillations (LCO) in a single oscillator (Rijke tube) shows the presence of Arnold tongue along with the region of asynchronous quenching for the parametric range of $f_f<f_{n0}$. The region of asynchronous quenching coincides with the boundary of forced synchronization of the acoustic pressure fluctuations in the system \cite{mondal2019forced}. The presence of Arnold tongue and asynchronous quenching are consistent with the results observed in previous studies conducted on Rijke tubes \cite{mondal2019forced} and laminar combustors \cite{guan2019open, roy2020mechanism, guan2019forced}. We notice that the characteristics of forced synchronization of LCO are dependent on the amplitude of LCO in the unforced state. In particular, the region of forced synchronization (or Arnold tongue) of the oscillator gets narrower as the amplitude of LCO is increased. In addition, the coupled response of two Rijke tube oscillators (A and B) show the occurrence of two different states of oscillation quenching, i.e., amplitude death and partial amplitude death. These states occur only when a finite frequency detuning is present between the oscillators, and the length of the coupling tube lies within a specific range.

To expand the parametric range of oscillation quenching in two mutually coupled Rijke tube oscillators, we acoustically force Rijke tube A. We observe that the suppression of LCO in coupled identical Rijke tube oscillators is possible through the combined effect of mutual coupling and asynchronous quenching. A significantly larger value of forcing amplitude is required to synchronize and quench the large amplitude LCO in the coupled thermoacoustic oscillators. This behavior is depicted through the increase in the steepness of the boundaries of Arnold tongue as the amplitude of LCO is increased. Suppression of LCO is observed predominantly for $f_f<f_{n0}$ in Rijke tube A, while it is observed on both sides of $f_{n0}$ in Rijke tube B. We notice that external forcing widens the region of coupling and forcing parameters over which the oscillation quenching states are observed in both the oscillators than when only one of the two mechanisms of control are applied on its own. Most importantly, although Rijke tube A is forced, the suppression of LCO is more significant in Rijke tube B in comparison with that in Rijke tube A. We also studied the coupled behavior of two non-identical thermoacoustic oscillators under asymmetric forcing. We witness that because of direct influence of forcing, Rijke tube A exhibits the features of forced synchronization, while Rijke tube B (not directly forced) remains desynchronized with the forcing signal. As a result, we observe a significant suppression of LCO in Rijke tube A and not in Rijke tube B, which is opposite to the behavior of forced response of coupled identical thermoacoustic oscillators. Finally, we qualitatively capture the experimental results through a reduced-order model of two coupled Rijke tubes. A good agreement is obtained between the experimental and numerical results.

Thus, periodic forcing aids the mitigation of thermoacoustic instability observed in coupled identical oscillators. We believe that the present investigation on the asymmetrically forced prototypical coupled thermoacoustic oscillators would prove to be a benchmark specifically for the control of thermoacoustic oscillations observed in can combustors with multiple cans, and for coupled nonlinear oscillators subjected to external forcing in general nonlinear dynamics literature. 

\section*{Acknowledgments}
A.S. and A.R. gratefully acknowledge the Ministry of Human Resource Development (MHRD) for funding PhD through Half-Time Research Assistantship (HTRA). The authors are grateful to Ms. Srikanth and Ms. Manoj for helping with the MATLAB implementation of the numerical model. This work was supported by the Office of Naval Research Global (Contract Monitor: Dr R. Kolar) Grant no. N62909-18-1-2061.

\bibliographystyle{apsrev4-2}
\bibliography{manuscript.bib}% Produces the bibliography via BibTeX.

\end{document}

\renewcommand{\theequation}{S\arabic{equation}}

\renewcommand{\thetable}{S\arabic{table}}  
\renewcommand{\thefigure}{S\arabic{figure}}

\preprint{APS/123-QED}

\title[]{\underline{Supplemental material} \\ Dynamics of coupled thermoacoustic oscillators\\ under asymmetric forcing: Experiments and theoretical modeling}

%\author[2]{Premchand C P}
%\author[1]{Manikandan Raghunathan}
%\author[1]{Abin Krishnan}
%\author[2]{Vineeth Nair}
%\author[1]{Sujith R I}
%\address[1]{Department of Aerospace Engineering, IIT Madras, Tamil Nadu - 600 036, India}
%\address[2]{Department of Aerospace Engineering, IIT Bombay, Maharashtra - 400 076, India}

\author{Ankit Sahay}
 \email[Corresponding Author: ]{ankitsahay02@gmail.com}
 \affiliation{Department of Aerospace Engineering, IIT Madras, Tamil Nadu - 600 036, India} 
\author{Amitesh Roy}
\affiliation{Department of Aerospace Engineering, IIT Madras, Tamil Nadu - 600 036, India}

\author{Samadhan A. Pawar}
\affiliation{Department of Aerospace Engineering, IIT Madras, Tamil Nadu - 600 036, India} 

\author{R I Sujith}
\affiliation{Department of Aerospace Engineering, IIT Madras, Tamil Nadu - 600 036, India} 

\date{\today}% It is always \today, today,

\maketitle

%\begin{quotation}
%This is the lead paragraph.
%\end{quotation}

\section{\label{R_square} $R^2$ values for synchronization boundaries}

\begin{table}[h]
\caption{\label{tab:table1} $R^2$ values of least-square-fitted boundaries of the Arnold tongue. The subscripts $l,r$ denote the left and right boundaries, respectively.}
\begin{ruledtabular}
\begin{tabular}{cccccc}
 & & \multicolumn{2}{c}{Rijke tube A}&\multicolumn{2}{c}{Rijke tube B}\\
 Figures & $p^\prime_{0}$ & $R^2_l$ & $R^2_r$ & $R^2_l$ & $R^2_r$ \\ \hline
 Fig. 2 & 120 Pa & 0.98 & 0.96 & - & - \\
        & 200 Pa & 0.96 & 0.85 & - & - \\
 Fig. 6 & 120 Pa & 0.99 & 0.88 & 0.96 & 1.00 \\
        & 200 Pa & 0.95 & 0.89 & 0.94 & 0.81 \\
 Fig. 7 & 120 Pa & 0.98 & 0.93 & 0.94 &  -   \\
        & 200 Pa & 0.97 & 0.99 &   -  &  -   \\
     Fig. 8 &        & 0.98 & 0.98 &  - & - \\
 Fig. 11 &       & 0.91 & 0.98 & 0.98 & 0.99 \\
 Fig. 12 &       & 0.98 & 0.98 & - & 0.98 \\
\end{tabular}
\end{ruledtabular}
\end{table}

Table \ref{tab:table1} shows the $R^2$ values obtained when a linear fit is applied to the data points on the forced synchronization boundaries in the $\bar{A}_f-f_f$ plane. Figures 2, 6 and 7 refer to experimental results shown in the main manuscript, whereas Figs. 8, 11 and 12 refer to results obtained from the mathematical model. An $R^2 = 1$ indicates that the linear regression predictions perfectly fit the data. 

\newpage

\section{Period-3 oscillations exhibit by a single Rijke tube oscillator under external forcing}
\label{Period_3}

\begin{figure*}[h!]
\centering
\includegraphics[scale=0.13]{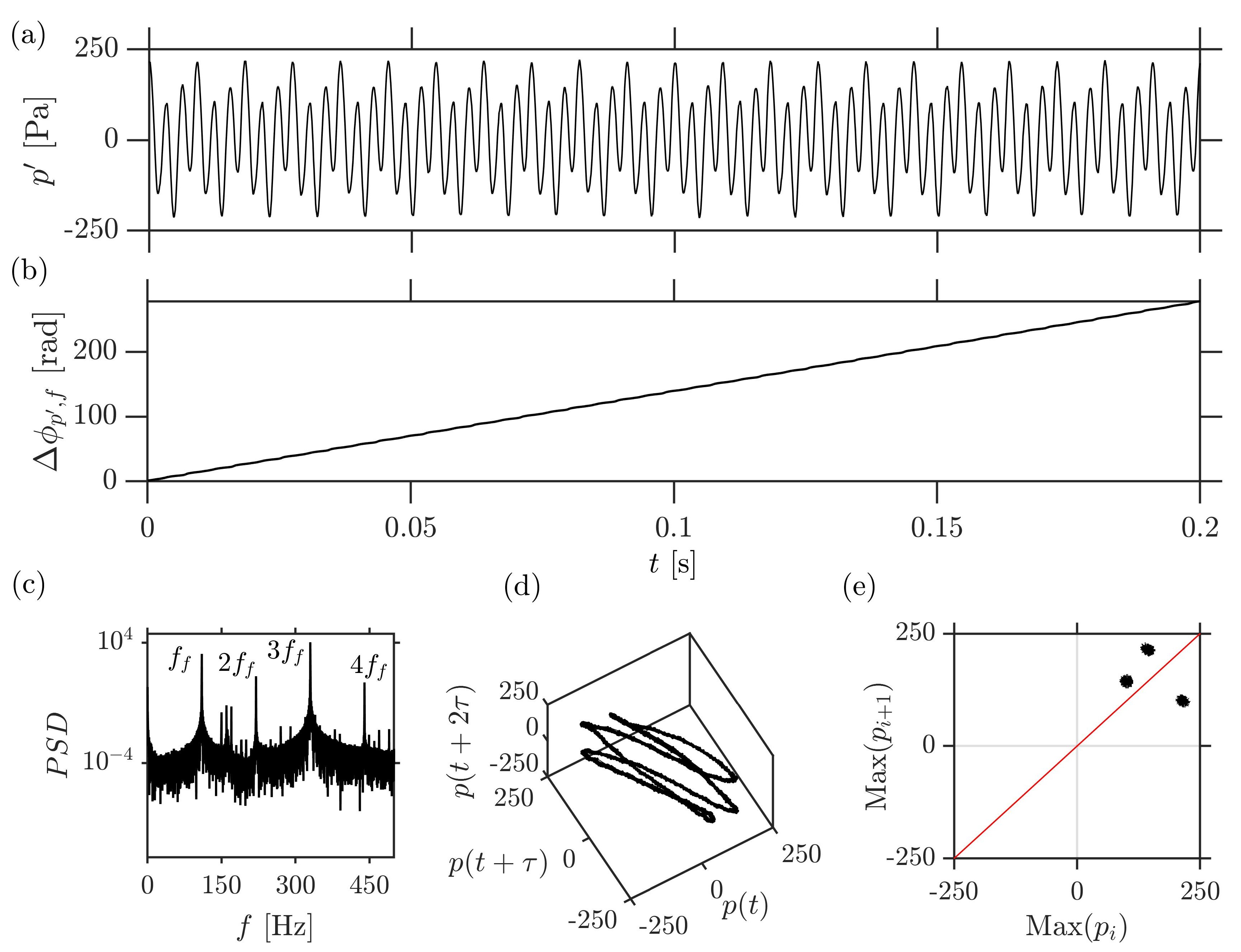}
\caption{\label{superharmonic} Time series of (a) the acoustic pressure fluctuations and (b) instantaneous phase difference between the pressure and the forcing signal. (c) The power spectrum, (d) the reconstructed phase portrait, and (e) the first return map of the forced acoustic pressure oscillations in a single Rijke tube exhibiting LCO of amplitude $p^\prime_{0} = 200$ Pa. The forcing is applied at $f_f/f_{n0} = 0.69$ and $\bar{A}_f= 0.65$. As a result, the acoustic pressure fluctuations show period-3 oscillations and, hence, remain desynchronized with the forcing signal, causing a lower value of \textit{PLV}. The period-3 oscillations are confirmed from the presence of three-looped attractor in the phase space and three fixed points in the return map.} 
\end{figure*}

During experiments in a single Rijke tube oscillator, we observe period-3 behavior in $p^\prime$ for high values of $\bar{A}_f$ in $f_f/f_{n0} = 0.62-0.70$ range, which leads to low $PLV$ calculated between the $p^\prime$ and forcing signal. In Fig. \ref{superharmonic}a, we show the acoustic pressure fluctuations exhibited by the Rijke tube at $f_f/f_{n0} = 0.69$ and $\bar{A}_f = 0.65$. The unforced amplitude of the LCO exhibited by the Rijke tube is $p^\prime_{0} = 200$ Pa. The period-3 behavior can be observed from the time series in Fig. \ref{superharmonic}a, as well as the spectral peaks in Fig. \ref{superharmonic}c, where we notice the presence of spectral peaks of approximately same magnitude at $f_f, 2f_f$ and $3f_f$ locations. Correspondingly, in Fig. \ref{superharmonic}d, the structure representative of the system dynamics (referred to hereinafter as the attractor) is a triple-looped attractor, i.e., the trajectories need to loop thrice before coming back to the initial point. Because the orbit is periodic, we get three distinct dots in the single-sided return map for the acoustic pressure time series in Fig. \ref{superharmonic}e.

\newpage

\section{ Effect of varying coupling tube parameters on the amplitude dynamics of an identical Rijke tubes coupled system}
\label{Supp_mat_3}

\begin{figure*}[h!]
\centering
\includegraphics[scale=0.45]{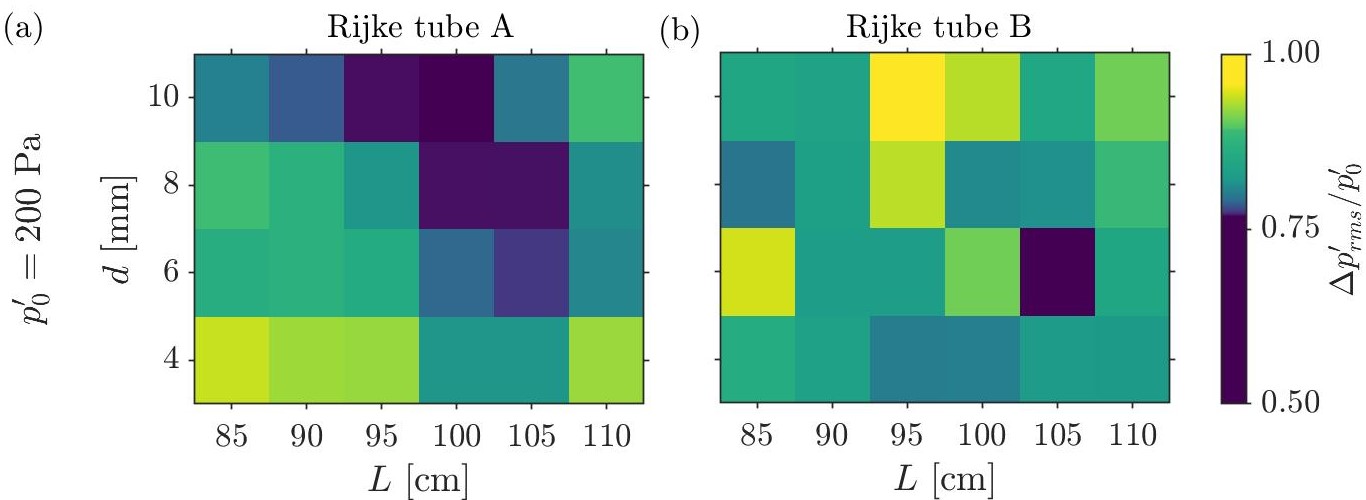}
\caption{\label{d_vs_l}Experimental two-parameter bifurcation plots showing the variation of $\Delta p^\prime_{rms}/p^\prime_{0}$ for different values of coupling tube length $L$ and internal diameter $d$ in identical Rijke tubes ($\Delta f_{n0} = 0$ Hz). The maximum suppression obtained is around 50\% in Rijke tube A for $L =100$ cm and $d =10$ mm.} 
\end{figure*}

Here, we explore experimentally the suppression of LCO in coupled Rijke tubes for connecting tubes of varying lengths and internal diameters. Figure \ref{d_vs_l} shows the reduction in the amplitude of LCO for different combinations of $L$ and $d$ for identical Rijke tubes. We notice that for $d=10$ mm, we obtain maximum suppression of about 50\% in the acoustic pressure fluctuations for $L \sim 100$ cm, in Rijke tube A. Thus, we keep $d=10$ mm for all our experiments. Large diameter connecting tubes may not be feasible for real-time combustors as it will require significant modification to engine hardware, whereas smaller diameters of connecting tube, such as the one used in the present study, can be very easily implemented.

\newpage

\section{ Mathematical model}
\label{model}

Here, we derive a reduced-order model for the coupled horizontal Rijke tubes subjected to asymmetric forcing. The model is based on \citet{balasubramanian2008thermoacoustic}. We neglect the effects of mean flow and mean temperature gradient in the duct. For two Rijke tubes A \& B, the acoustic momentum and energy equations for a medium with a perfect, inviscid and non-heat conducting gas are then given as \cite{balasubramanian2008thermoacoustic}:
\begin{equation} \label{eq:1}
    \bar{\rho}\dfrac{\partial \Tilde{u_a}^\prime}{\partial \Tilde{t}} + \dfrac{\partial \Tilde{p_a}^\prime}{\partial \Tilde{x}} = 0,
\end{equation}

\begin{equation} \label{eq:2}
    \dfrac{\partial \Tilde{p_a}^\prime}{\partial \Tilde{t}} + \gamma \bar{p}\dfrac{\partial \Tilde{u_a}^\prime}{\partial \Tilde{x}} = (\gamma -1)\Dot{\Tilde{Q}}^\prime \delta(\Tilde{x}-\Tilde{x_f}),
\end{equation}

\begin{equation} \label{eq:3}
    \bar{\rho}\dfrac{\partial \Tilde{u_b}^\prime}{\partial \Tilde{t}} + \dfrac{\partial \Tilde{p_b}^\prime}{\partial \Tilde{x}} = 0,
\end{equation}

\begin{equation} \label{eq:4}
    \dfrac{\partial \Tilde{p_b}^\prime}{\partial \Tilde{t}} + \gamma \bar{p}\dfrac{\partial \Tilde{u_b}^\prime}{\partial \Tilde{x}} = (\gamma -1)\Dot{\Tilde{Q}}^\prime \delta(\Tilde{x}-\Tilde{x_f}).
\end{equation}

where, $\Tilde{p}^\prime$ and $\Tilde{u}^\prime$ are the acoustic pressure and velocity fluctuations, respectively. $\gamma$ is the ratio of specific heats of air at ambient conditions. $\Tilde{x}$ is the distance along the axial direction in the duct, $\Tilde{t}$ is the time, $\bar{\rho}$ and $\bar{p}$ are the ambient density and pressure, respectively. Subscripts $a$ and $b$ indicate the quantities correpsonding to Rijke tube A \& B, respectively. For a general system of non-identical oscillators, we define a quantity $r$ as:
\begin{equation} \label{eq:5}
r = \dfrac{L_b}{L_a} = \dfrac{\omega_b}{\omega_a},
\end{equation}
where, $L_a$ and $L_b$ are lengths of the Rije tube ducts A \& B, respectively. $\omega_a$ and $\omega_b$ are the natural frequencies of the Rijke tubes A and B, respectively.

We non-dimensionalize Eqs. (\ref{eq:1}) and (\ref{eq:4}) using the following transformations:
\begin{equation} \label{eq:6}
    x = \dfrac{\Tilde{x}}{L_a}; \hspace{5 pt} t = \dfrac{c_0}{L_a}\Tilde{t}; \hspace{5 pt} u_{a}^\prime = \dfrac{\Tilde{u_a}^\prime}{u_0}; \hspace{5 pt} p_{a}^\prime = \dfrac{\Tilde{p_a}^\prime}{\bar{p}}; \hspace{5 pt} M = \dfrac{u_0}{c_0}; \hspace{5 pt} u_{b}^\prime = \dfrac{\Tilde{u_b}^\prime}{u_0}; \hspace{5 pt} p_{b}^\prime = \dfrac{\Tilde{p_b}^\prime}{\bar{p}}; \hspace{5 pt} \Dot{Q}^\prime = \dfrac{\Dot{\Tilde{Q}}^\prime}{c_0 \bar{p}}.
\end{equation}
Here, variables with tilde are dimensional and variables without tilde are non-dimensional. $u_0$ and $\bar{p}$ are the steady state velocity and pressure of the flow, respectively. $c_0$ is the speed of sound, and $M$ is the Mach number of the mean flow. $x$ and $t$ are the non-dimensional axial distance and time, respectively. Using the above transformations, we obtain the following non-dimensionlized acoustic momentum and energy equations:
\begin{equation} \label{eq:7}
    \gamma M \dfrac{\partial u_a^\prime}{\partial t} + \dfrac{\partial p_a^\prime}{\partial x} = 0,
\end{equation}

\begin{equation} \label{eq:8}
    \dfrac{\partial p_a^\prime}{\partial t} + \gamma M \dfrac{\partial u_a^\prime}{\partial x} = \dfrac{(\gamma -1)L_a \Dot{Q}^\prime}{\bar{p}c_0} \delta[L_a(x-x_f)],
\end{equation}

\begin{equation} \label{eq:9}
\gamma M \dfrac{\partial u_b^\prime}{\partial t} + \dfrac{\partial p_b^\prime}{\partial x} = 0,
\end{equation}

\begin{equation} \label{eq:10}
\dfrac{\partial p_b^\prime}{\partial t} + \gamma M \dfrac{\partial u_b^\prime}{\partial x} = \dfrac{(\gamma -1)L_a \Dot{Q}^\prime}{\bar{p}c_0} \delta[L_a(x-x_f)].
\end{equation}

The heat release rate $\Dot{Q}^\prime$ is modeled using a modified form of King's law \cite{king1914xii, heckl1990non} which correlates the quasi-steady heat transfer from a heated cylinder to the flow around it. The expression for normalized heat release rate fluctuations is written in terms of the acoustic velocity fluctuations, observed at the heater location $x_f$ after time delay $\tau$ as:
\begin{equation} \label{eq:11}
    \Dot{Q}^\prime = \dfrac{2L_{w}(T_{w}-\bar{T})}{S\sqrt{3}} \sqrt{\pi \lambda C_{v}\bar{\rho} \dfrac{d_w}{2}}
 \left[  \sqrt{\left| \dfrac{u_0}{3} + u_{f}^\prime(t-\tau) \right|}-\sqrt{\dfrac{u_0}{3}}\right],
\end{equation} where, $d_w$, $L_w$ and $T_w$ are the diameter, length and temperature of the heater wire, respectively. $\bar{T}$ is the steady state temperature of the flow, $S$ is the cross-sectional area of the duct, $C_v$ \& $\lambda$ are the specific heat at constant volume and thermal conductivity, respectively, of the medium within the duct. $\tau$ quantifies the thermal inertia of the heat transfer from the heating element to the medium.

The non-dimensionalized set of PDEs in Eqs. (\ref{eq:7})-(\ref{eq:10}) is reduced to a set of ordinary differential equations using the Galerkin technique \cite{lores1973nonlinear}. To that end, the non-dimensional velocity $u^\prime$ and non-dimensional pressure $p^\prime$ fluctuations in the model are written in terms of the Galerkin modes:

\begin{equation} \label{eq:12}
    p_a^\prime(x,t) =  \sum_{j=1}^{N} -\dfrac{\gamma M}{j \pi}\dot{\eta^a_{j}}(t)\hspace{2 pt}\text{sin}(j \pi x),
\end{equation}

\begin{equation} \label{eq:13}
    u_a^\prime(x,t) =  \sum_{j=1}^{N} \eta^a_{j}(t)\hspace{2 pt}\text{cos}(j \pi x),
\end{equation}

\begin{equation} \label{eq:14}
p_b^\prime(x,t) =  \sum_{j=1}^{N} -\dfrac{\gamma M r}{j \pi}\dot{\eta^b_{j}}(t)\hspace{2 pt}\text{sin}\left(\dfrac{j \pi x}{r}\right),
\end{equation}

\begin{equation} \label{eq:15}
    u_b^\prime(x,t) =  \sum_{j=1}^{N} \eta^b_{j}(t)\hspace{2 pt}\text{cos}\left(\dfrac{j \pi x}{r}\right),
\end{equation}
Here, $\eta_j$ and $\dot{\eta_j}$ represent the time-varying coefficients of the $j$th mode of the acoustic velocity $u^\prime$ and acoustic pressure $p^\prime$, respectively. $a$ and $b$ correspond to the acoustic variables in Rijke  tubes A and B, respectively. We can verify that the particular form of Galerkin modes satisfies the acoustically open-open boundary conditions: $p_a^\prime (x=0,t)=0$, $p_a^\prime (x=1,t)=0$, $p_b^\prime (x=0,t)=0$ and $p_b^\prime (x=r,t)=0$ . $N$ represents the number of Galerkin modes considered. Substituting Eqs. (\ref{eq:12})-(\ref{eq:15}) in Eqs. (\ref{eq:7})-(\ref{eq:10}) with a damping term included \cite{matveev2003thermoacoustic}, and projecting the resultant equations along the basis functions, we obtain the following set of first-order ordinary differential equations:
\begin{equation}  \label{eq:16}
    \dfrac{d \eta^a_j}{dt} = \Dot{\eta^a_{j}},
\end{equation}
\begin{equation}  \label{eq:17}
    \dfrac{d \Dot{\eta^a_{j}}}{dt} + 2 \xi_j \omega_j \Dot{\eta^a_{j}} + k_{j}^2 \eta^a_{j} = -2 j \pi K \left[\sqrt{\left| \dfrac{1}{3} + u_{f}^{\prime a}(t-\tau) \right|} -\sqrt{\dfrac{1}{3}}\right]\hspace{2 pt}\text{sin}(j\pi x_f),
\end{equation}

\begin{equation}  \label{eq:18}
    \dfrac{d \eta^b_j}{dt} = \Dot{\eta^b_{j}},
\end{equation}
\begin{equation}  \label{eq:19}
    \dfrac{d \Dot{\eta^b_{j}}}{dt} + 2 \xi_j \left(\dfrac{\omega_j}{r}\right) \Dot{\eta^b_{j}} + \left(\dfrac{k_{j}}{r}\right)^2 \eta^b_{j} = -\dfrac{2 j \pi K}{r^2} \left[\sqrt{\left| \dfrac{1}{3} + u_{f}^{\prime b}(t-\tau) \right|} -\sqrt{\dfrac{1}{3}}\right]\hspace{2 pt}\text{sin}\left(\dfrac{j\pi x_f}{r}\right),
\end{equation}

 where, $k_j = j\pi$ refers to the non-dimensional wave number and $\omega_j = j \pi$ refers to the non-dimensional angular frequency of the $j$th mode. The coefficient $\xi_j$ appearing in the second term of Eqs. (\ref{eq:17}) \&   (\ref{eq:19}) represents the frequency-dependent damping \cite{matveev2003thermoacoustic}, and is given by the following ansatz \cite{sterling1991nonlinear}:

\begin{equation} \label{eq:20}
    \xi_j = \dfrac{c_1 \dfrac{\omega_j}{\omega_1}+c_2\sqrt{\dfrac{\omega_1}{\omega_j}}}{2\pi}
\end{equation}
Here, $c_1$ and $c_2$ are the damping coefficients which determine the amount of damping. We choose the values $c_1=0.1$ and $c_2=0.06$ based on \cite{sterling1991nonlinear} for all simulations.

To formulate a model for the coupled system, we assume that the two Rijke tubes are coupled through time-delay and dissipative couplings. Based on Eqs. (\ref{eq:16})-(\ref{eq:19}), the governing equations for non-identical, coupled Rijke tubes with asymmetric sinusoidal forcing can be written as:

\begin{equation} \label{eq:21}
    \dfrac{d \eta_j^{a,b}}{dt} = \Dot{\eta_{j}^{a,b}},
\end{equation}
\begin{equation} \label{eq:22}
\begin{split}
    \dfrac{d \Dot{\eta_{j}}^{a,b}}{dt} + 2 \xi_j \left(\dfrac{\omega_j}{r^{a,b}}\right) \Dot{\eta_{j}^{a,b}} + \left(\dfrac{k_{j}}{r^{a,b}}\right)^2 \eta_{j}^{a,b} = -\dfrac{2 j \pi K}{{r^{a,b}}^2} \left[\sqrt{\left| \dfrac{1}{3} + u_{f}^{\prime a,b} (t-\tau) \right|} -\sqrt{\dfrac{1}{3}}\right]\hspace{2 pt}\text{sin}\left(\dfrac{j\pi x_f}{r^{a,b}}\right) \\ + \underbrace{K_d(\dot{\eta_j}^b - \dot{\eta_j}^a)}_\text{Dissipative coupling} + \underbrace{K_\tau (\dot{\eta_j}^b(t-\tau_{tube})-\dot{\eta_j}^a(t))}_\text{Time-delay coupling} + \underbrace{[A_f \hspace{2 pt} \text{sin}(2 \pi f_f t)]^a}_\text{Forcing term},
\end{split}
\end{equation} 
where, $r^{a,b}$ is defined as the ratio of the length of the duct to a reference length, $L_{a,b}/L_{ref}$. We consider $L_a$ to be the reference length. For identical oscillators, $r^a=r^b = 1$. For non-identical oscillators, $r^a = 1$ and $r^b = L_b/L_a = \omega_a/\omega_b$.

The second and third terms on the right-hand side of Eq. (\ref{eq:14}) are the dissipative and time-delay coupling terms, respectively. Dissipative coupling encapsulates the interaction that arises from the mass transfer between the two ducts \cite{bar1985stability}. Time-delay coupling quantifies the time taken by acoustic waves to propagate through the coupling tube connecting the two Rijke tubes. Thus, $\tau_{tube}$ denotes the time-delay in the response induced by one Rijke tube on the other. The fourth term is the sinusoidal forcing term with amplitude $A_f$ and frequency $f_f$. The external forcing is applied only to Rijke tube A.

\bibliographystyle{apsrev4-2}
\bibliography{supplemental.bib}% Produces the bibliography via BibTeX.